\documentclass[10pt]{article}

\usepackage[margin=1in]{geometry}
\usepackage{titlesec}
\usepackage{subcaption}
\usepackage[labelfont=bf,textfont=it,font=small]{caption}
\usepackage{parskip}
\usepackage[comma,sort&compress,numbers]{natbib}
\usepackage{enumitem}
\usepackage{bibunits}
\usepackage{setspace}
\usepackage{enumitem}

\usepackage{amssymb}
\usepackage{amsthm}
\usepackage{amsmath}
\usepackage{amsfonts}       
\usepackage[ruled]{algorithm2e}
\usepackage{array}
\usepackage{graphicx}
\usepackage{tabulary}
\usepackage{multirow}
\usepackage{todonotes}
\usepackage{url}
\usepackage{nicefrac}       
\usepackage{microtype}      
\usepackage{siunitx}
\usepackage{tikz,pgfplots}
\usetikzlibrary{decorations,
                arrows,
                shapes,
                trees,
                shapes.geometric,
                calc,
                arrows.meta,
                positioning,
                fit, 
                shapes.multipart,
                backgrounds}
\usepackage{fontawesome5}
\pgfplotsset{compat=1.16}
\usepackage{adjustbox}

\usepackage[utf8]{inputenc} 
\usepackage[T1]{fontenc}    
\usepackage{times}
\usepackage{helvet}
\usepackage[pagewise]{lineno} 
\usepackage{booktabs}

\usepackage{xcolor}         
\usepackage{svg}

\usepackage{hyperref}
\hypersetup{colorlinks,
      linkcolor=blue,
      citecolor=blue,
      urlcolor=blue}

\newcommand{\cref}[2]{\hyperref[#2]{#1~\ref*{#2}}}
\newcommand{\colref}[3]{\hyperref[#2]{#1~\ref*{#2}{#3}}}
\newcommand{\figref}[1]{\cref{Figure}{#1}}

\newcommand{\figrefs}[2]{\hyperref[#1]{Figure~(\ref*{#1}--\ref*{#2})}}

\titleformat{\section}{\large\bfseries\sffamily}{\thesection}{1em}{}
\titleformat{\subsection}{\normalsize\bfseries\sffamily}{\thesubsection}{1em}{}
\titleformat{\subsubsection}{\small\sffamily\bfseries}{\thesubsubsection}{1em}{}

\title{3D Multiphase Heterogeneous Microstructure Generation Using Conditional Latent Diffusion Models}

\author{Nirmal Baishnab, Ethan Herron, Aditya Balu, Soumik Sarkar, \\Adarsh Krishnamurthy$^*$, Baskar Ganapathysubramanian$^*$\\
\small Iowa State University, Ames, IA, USA\\ \small $^*$ Corresponding authors: (adarsh | baskar)@iastate.edu}

\date{}

\begin{document}
\maketitle
\begin{abstract}
The ability to generate 3D multiphase microstructures on-demand with targeted attributes can greatly accelerate the design of advanced materials. Here, we present a conditional latent diffusion model (LDM) framework that rapidly synthesizes high-fidelity 3D multiphase microstructures tailored to user specifications. Using this approach, we generate diverse two-phase and three-phase microstructures at high resolution (volumes of $128 \times 128 \times 64$ voxels, representing $>10^6$ voxels each) within seconds, overcoming the scalability and time limitations of traditional simulation-based methods. Key design features, such as desired volume fractions and tortuosities, are incorporated as controllable inputs to guide the generative process, ensuring that the output structures meet prescribed statistical and topological targets. Moreover, the framework predicts corresponding manufacturing (processing) parameters for each generated microstructure, helping to bridge the gap between digital microstructure design and experimental fabrication. While demonstrated on organic photovoltaic (OPV) active-layer morphologies, the flexible architecture of our approach makes it readily adaptable to other material systems and microstructure datasets. By combining computational efficiency, adaptability, and experimental relevance, this framework addresses major limitations of existing methods and offers a powerful tool for accelerated materials discovery.
\end{abstract}

\section{Introduction}
Understanding and controlling a material’s microstructure is critical for optimizing its properties and performance. In materials science, the mapping between structure and property is a foundational concept, with microstructural features often serving as primary drivers of a material’s physical characteristics and behavior~\citep{newnham2012structure,le2012quantitative,carraher2012structure,li2020towards}. However, directly observing or reconstructing 3D microstructures through experiments is expensive and technically challenging, making it difficult to explore processing–structure–property relationships at scale~\citep{midgley2009electron, scott2012electron, franken2017transmission,  mohammed2018scanning}. Consequently, there is a strong motivation to develop computational methods for generating realistic microstructures. The ability to produce statistically representative microstructure samples on-demand would greatly aid in virtual testing, microstructure-sensitive property prediction, and computational materials design.

Various approaches have been explored for microstructure generation~\citep{bostanabad2018computational, torquato2002random}. Classical statistical methods, such as Markov random fields~\citep{bostanabad2016stochastic}, Gaussian random fields~\citep{jiang2013efficient}, and descriptor-based reconstructions~\citep{xu2014descriptor,jiao2008modeling}, can produce microstructures that match certain target statistics. While these methods have proven useful, they suffer from important limitations. In general, statistical models are computationally intensive and do not scale well to generating large 3D volumes or numerous samples. They often rely on strict assumptions (e.g. stationarity or isotropy of features) and tailored mathematical descriptors, which limits their flexibility and generalizability to different materials or complex structures. Adapting such models to incorporate new microstructural constraints or application-specific objectives is non-trivial and typically requires substantial rederivation or optimization changes. These challenges highlight the need for a more flexible, data-driven generative framework for microstructures.

Recently, deep generative models have shown great promise in capturing complex microstructural features from data~\citep{bandi2023power, cao2023comprehensive}. Approaches like variational autoencoders (VAEs)~\citep{kingma2013auto}, generative adversarial networks (GANs)~\citep{goodfellow2020generative}, and diffusion models (DMs)~\citep{sohl2015deep} have been applied to microstructure generation tasks. VAEs can learn low-dimensional representations of microstructures but often produce blurry outputs that lack sharp detail ~\citep{wang2021generative}. GAN-based models have succeeded in generating 3D microstructures with improved visual fidelity~\citep{henkes2022three, hsu2021microstructure, chun2020deep}, but they do not allow user control over generated structures and are notorious for training instabilities~\citep{salimans2016improved}. Moreover, GANs and similar networks can be computationally demanding for 3D data, sometimes requiring extensive resources for training and generation. Diffusion models offer even higher output quality, often surpassing GANs, but their iterative sampling process makes inference slow and resource-intensive\citep{dhariwal2021diffusion}. At this time, no prior generative approach has simultaneously provided high fidelity, user controllability, and computational efficiency for 3D microstructure generation.

Latent diffusion models (LDMs) have emerged as a compelling solution to address these gaps~\citep{rombach2022high, croitoru2023diffusion}. LDMs combine the strengths of VAEs and DMs by operating in a compressed latent space to dramatically reduce computational costs while preserving the ability to generate high-quality, diverse microstructures. This latent-space approach yields orders-of-magnitude speed-ups over conventional pixel-space diffusion models. Importantly, LDM architectures naturally support conditioning mechanisms that enable users to steer generation towards desired attributes. They also exhibit more stable training dynamics and avoid mode collapse, yielding a broader variety of outputs compared to GANs~\citep{du2024confild, herron2023latent, pinaya2022brain, blattmann2023align}. These advantages make LDMs well-suited for fast and controllable 3D microstructure synthesis. 

To date, applying diffusion-based generative models to microstructure design has predominantly focused on unconditional generation~\citep{lee2024microstructure, lyu2024microstructure, fernandez2024digital}. In our prior work, Herron et al.~\citep{herron2022generative} applied a diffusion model to 2D organic solar cell microstructures without enabling user-specified target features. While recent advances~\citep{gao2025advanced, lee2024denoising} have begun exploring conditional generative approaches to microstructure reconstruction and design, these have typically not integrated predictions of corresponding manufacturing parameters. Our current work introduces a conditional latent diffusion modeling (LDM) framework that not only allows user-defined control over critical microstructural descriptors but also uniquely predicts manufacturing parameters likely to produce such microstructures experimentally. This two-fold capability addresses key challenges in computational materials design~\citep{kuehmann2009computational, panchal2013key}: not only can we generate microstructures with tailored properties, but we can also provide insight into how to manufacture them – thereby tackling the oft-cited “\textit{manufacturability gap}” in microstructure design.

We demonstrate the framework using organic photovoltaic (OPV) active-layer microstructures as a representative example. OPV active layers typically consist of a donor material and an acceptor material, forming a complex two-phase (or three-phase with a mixed phase) morphology~\citep{lee2010structural}. Two microstructural descriptors are particularly crucial for OPV performance: the donor (acceptor) phase volume fraction and the tortuosity of the percolating pathways~\citep{liu2012morphology, heiber2017impact}. The volume fraction (the ratio of donor to acceptor material in the blend) directly influences the balance between charge generation and transport, while tortuosity reflects the complexity of pathways that charge carriers must navigate to reach the electrodes. By conditioning on these properties in the LDM, we can generate microstructures that meet specific targets (e.g. a desired donor volume fraction and phase connectivity) known to optimize OPV efficiency. We quantify volume fraction and tortuosity for each generated sample using established computational techniques~\citep{wodo2013quantifying}.

The key contributions of this work include: (1) \textbf{\textit{Scalable high-resolution 3D microstructure generation}}: Leveraging an LDM, we rapidly produce diverse multiphase 3D microstructures (including two-phase and three-phase examples) at a resolution of $128 \times 128 \times 64$ voxels (over one million voxels each), which is orders of magnitude larger than those demonstrated in prior studies. Our approach generates these 3D microstructures in seconds per sample (versus hours or days with physics-based simulations). (2) \textbf{\textit{Conditional generation with user-defined features}}: Our framework introduces controllability to microstructure synthesis by allowing users to specify target volume fractions and tortuosities; the LDM then generates microstructures that faithfully realize these input parameters, ensuring the output matches desired structural characteristics. (3) \textbf{\textit{Linking microstructure to manufacturing}}: We integrate a predictive module that outputs relevant processing parameters (e.g. annealing or fabrication conditions) corresponding to each generated microstructure, facilitating a direct connection between the digital microstructure design and its experimental realization. These advances collectively overcome the scalability, controllability, and manufacturability limitations of existing methods. By enabling fast generation of application-specific microstructures along with guidance for their fabrication, our conditional LDM framework illustrates the promise of AI-driven approaches in computational materials science and microstructure design.

\pagebreak

\section{Results and Discussion}
To demonstrate our framework’s capabilities, we evaluated its performance using both synthetic microstructures generated via physics-based simulations (Cahn–Hilliard equation) and experimentally obtained (via tomography) organic photovoltaic (OPV) morphologies. The results illustrate the advantages of our conditional latent diffusion modeling (LDM) approach in generating diverse, high-quality microstructures efficiently and with precision.

Our proposed generative modeling framework, schematically illustrated in \figref{fig:Training}, consists of three sequentially trained modules: a Variational Autoencoder (VAE), a Feature Predictor (FP), and the Latent Diffusion Model (LDM). Initially, the VAE compresses complex, high-dimensional 3D microstructures into compact latent representations, drastically reducing computational complexity. The FP network subsequently predicts relevant microstructural features (e.g., volume fractions and tortuosities) and manufacturing parameters directly from these latent representations. Finally, the conditional LDM leverages these predictions to generate realistic 3D microstructures, guided explicitly by user-specified conditions.

In the following sub-sections, we detail our evaluation of the framework’s generative capabilities, including the quality and diversity of generated microstructures, the effectiveness of conditional sampling for targeted microstructure design, and the model's unique capacity to predict experimental manufacturing parameters. 

\subsection{Sampling quality}\label{sec:sampling_quality}
\begin{figure}[!b]
    \centering
    \begin{subfigure}[b]{0.45\linewidth}
        \includegraphics[width=\linewidth]{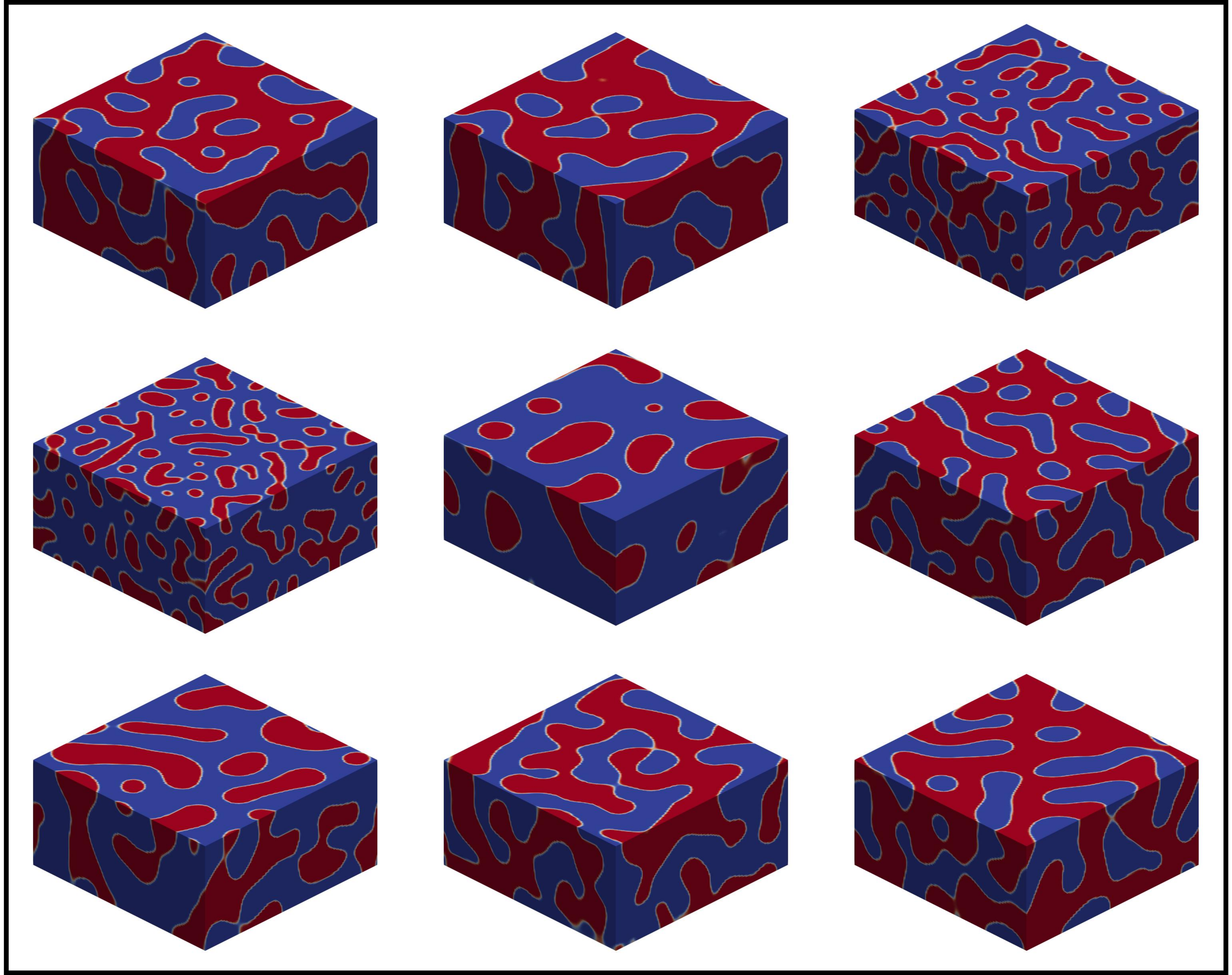}
        \caption{Two phase}
        \label{fig:two_phase}
    \end{subfigure}
    \hfill
    \begin{subfigure}[b]{0.45\linewidth}
        \includegraphics[width=\linewidth]{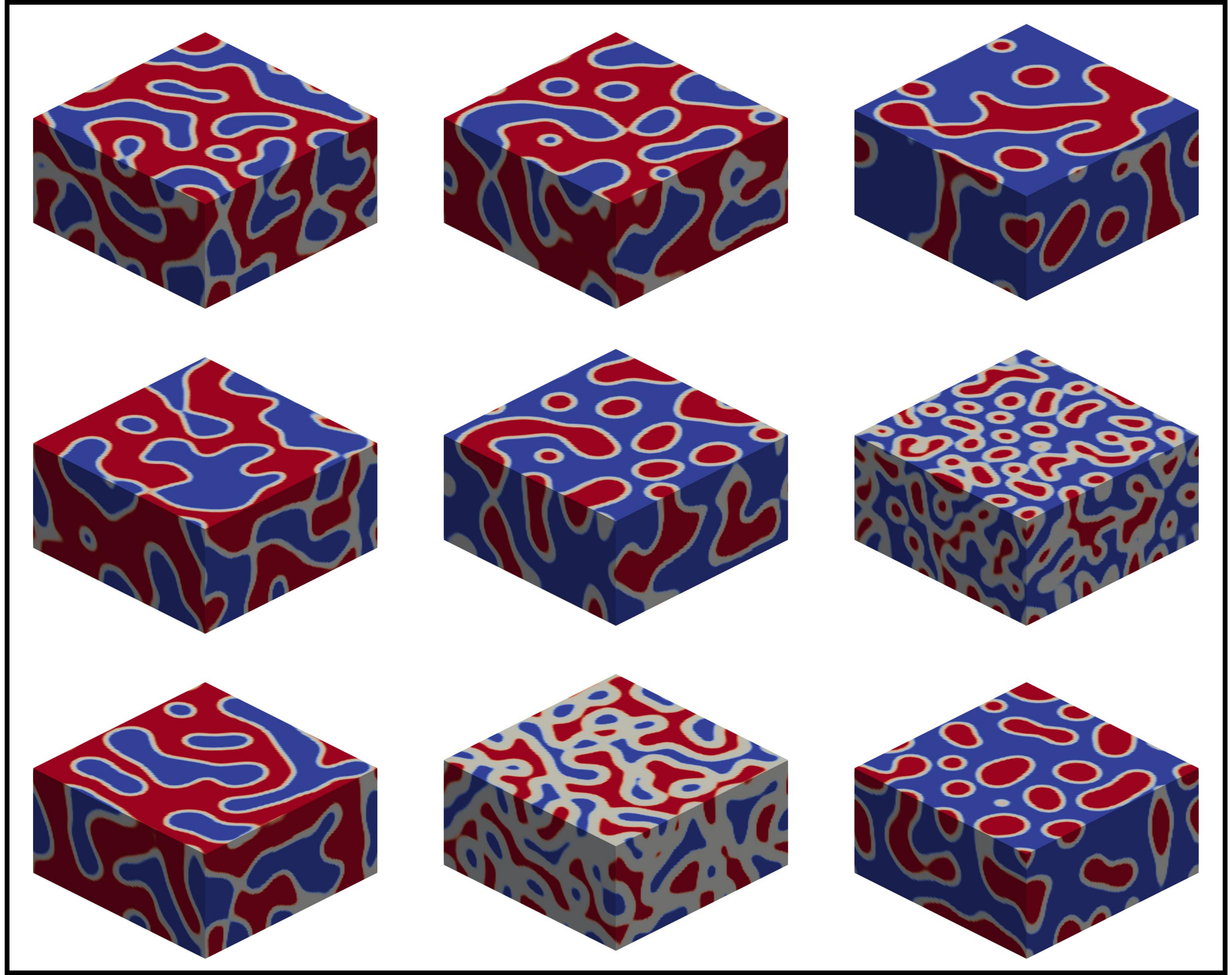}
        \caption{Three phase}
        \label{fig:three_phase}
    \end{subfigure}
    \caption{Samples from LDMs trained on (a) two phase and (b) three phase microstructures.}
    \label{fig:general_samples}
\end{figure}

\figref{fig:general_samples} shows representative examples of microstructures generated by our Latent Diffusion Models (LDMs), separately trained for two-phase and three-phase systems. In the two-phase microstructures, the blue domains represent a donor phase (denoted phase A), and the red domains represent an acceptor phase (denoted phase B), corresponding to typical organic photovoltaic (OPV) active-layer morphologies. For the three-phase microstructures, an additional gray phase delineates a mixed region, that typically exists as an interfacial region between donor and acceptor phases.

Each generated microstructure spans a volume of $128 \times 128 \times 64$ voxels, corresponding to over one million voxels (1,048,576), allowing detailed resolution of intricate morphological features. Importantly, our LDM framework achieves this generation within approximately 0.5 seconds per microstructure using an NVIDIA A100 GPU, significantly outperforming traditional physics-based simulation methods, which typically require hours or days of computation for similar-sized volumes \citep{wodo2011computationally, vondrous2014parallel, li2017computationally}.

The transition from two-phase to three-phase systems maintains high quality and fidelity, demonstrating the flexibility and scalability of our framework. Without any modification to the core architecture, retraining on a three-phase dataset successfully generated microstructures exhibiting smaller domains and more complex, finely detailed features. This ease of adaptability underscores the potential for further extension of our approach to accommodate additional phases.

\subsection{Conditional sampling}\label{sec:conditional_sampling}

In this work, conditional sampling refers to the approach of providing the generative model with additional information—termed a conditioning vector—to guide the synthesis of microstructures toward specific, user-defined characteristics. We implemented this conditional generation by embedding the conditioning vector directly into the latent diffusion model (LDM), allowing precise control over the structural features of the generated microstructures. Specifically, the LDM architecture incorporates the conditioning vector into the embedding layers of the U-Net backbone, facilitating effective guidance during the diffusion process (details available in Supplementary Information).

The LDM is conditioned on two crucial microstructural descriptors relevant to organic photovoltaics: the volume fractions and tortuosities of the phases (A, B, and the mixed phase). However, our flexible conditioning framework is easily extensible to other relevant morphological descriptors, depending on the application requirements (see additional examples provided in the Supplementary Results). \figref{fig:conditional_generation_3p_samples} illustrates representative examples of conditionally generated microstructures, clearly demonstrating the effectiveness of the model in synthesizing morphologies tailored to user-specified volume fractions and tortuosities.

\begin{figure}[!b]
    \centering
    \begin{subfigure}[b]{0.48\linewidth}
        \centering
        \includegraphics[width=0.99\linewidth]{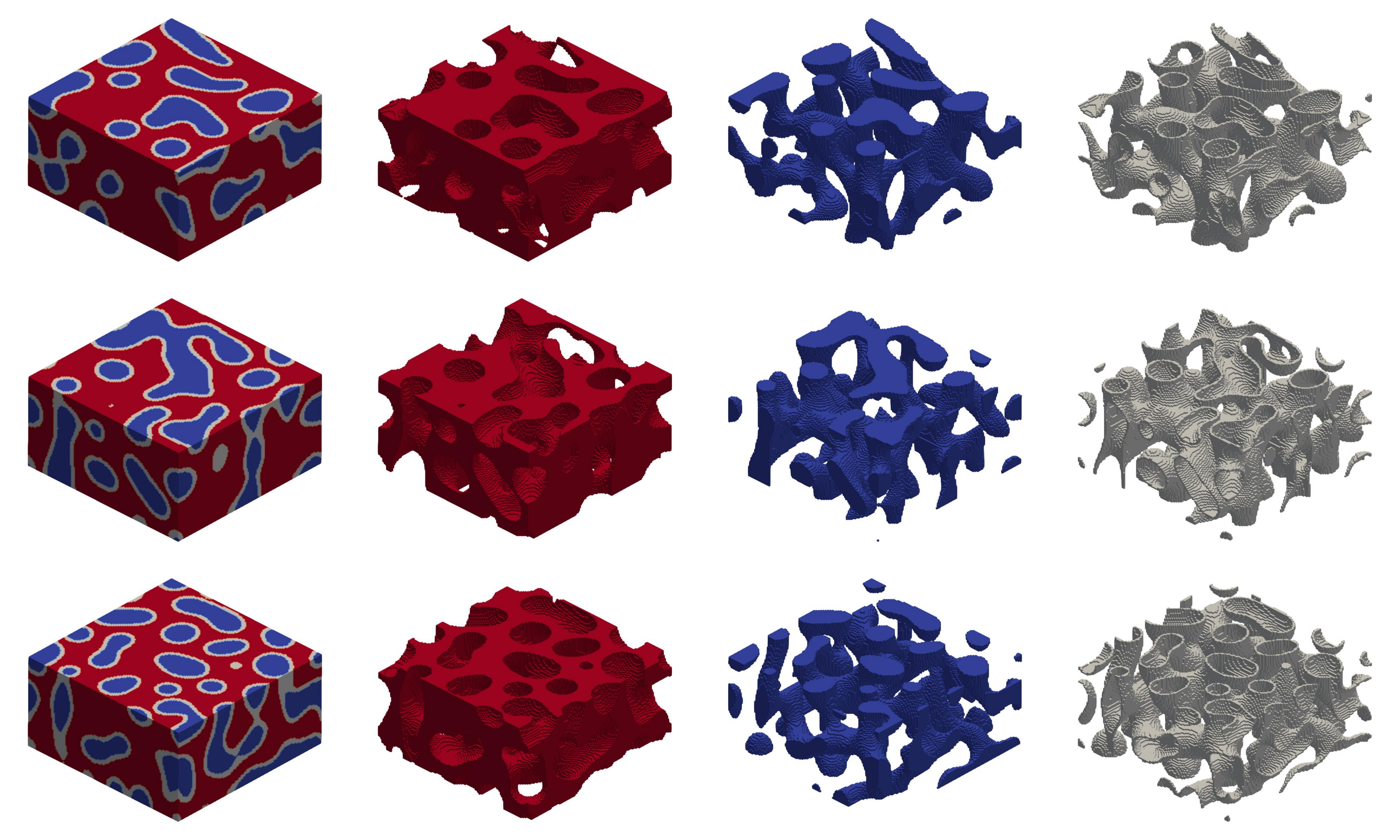}
        \subcaption{Predominant phase A - more than 0.5}
        \label{fig:phaseD}
    \end{subfigure}
    \begin{subfigure}[b]{0.48\linewidth}
        \centering
        \includegraphics[width=0.99\linewidth]{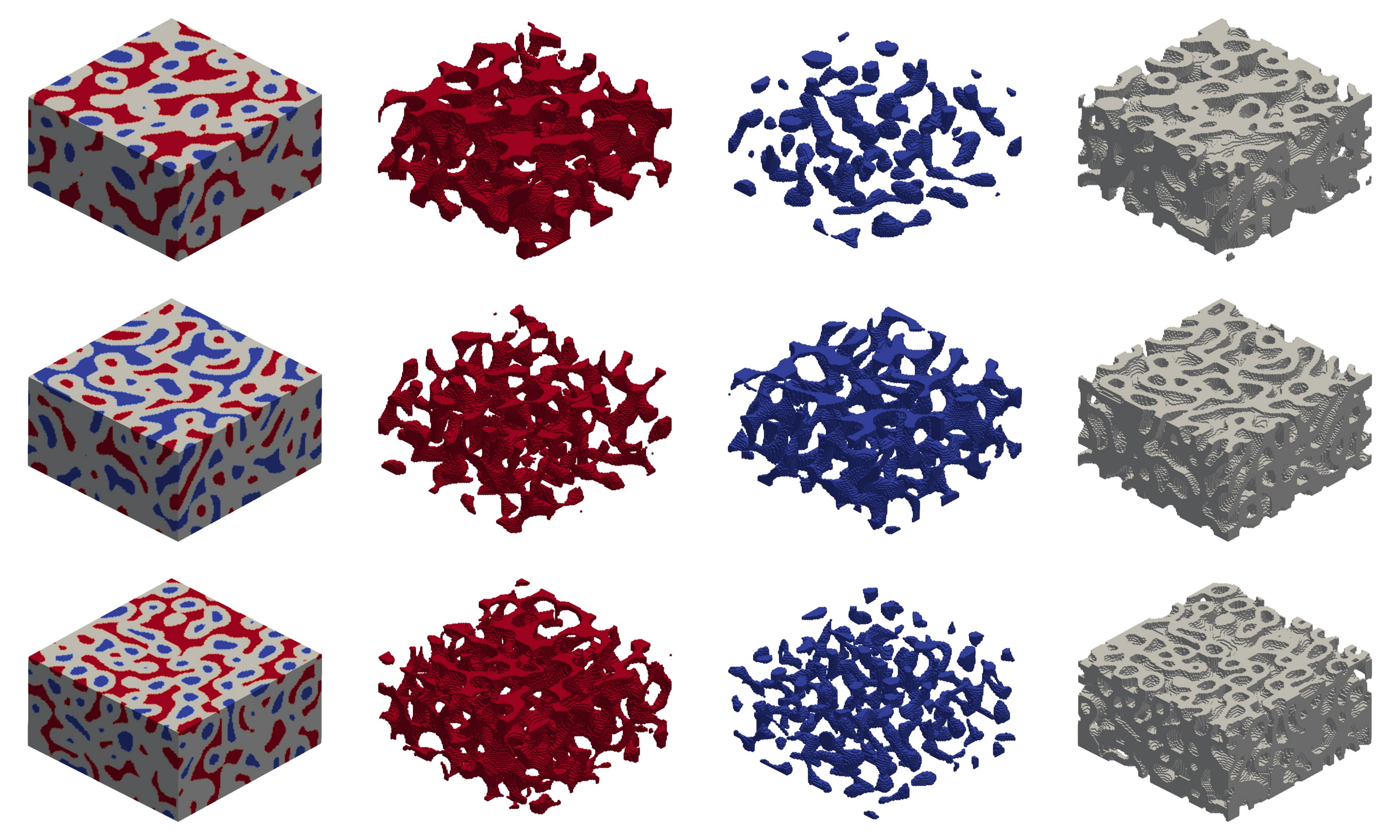}
        \subcaption{Predominant phase mixed - more than 0.5}
        \label{fig:phaseC}
    \end{subfigure}
    \caption{Conditional microstructure generation: Sample microstructures from user inputs - (a)~Predominant phase A, and (b)~Predominant phase mixed. First column shows the total microstructure. Second, third and fourth columns show the thresholded versions of the phase A, phase B  and mixed components, respectively.}
    \label{fig:conditional_generation_3p_samples}
\end{figure}

\begin{figure}[b!]
    \centering
    \begin{subfigure}[b]{0.19\linewidth}
        \includegraphics[width=0.99\linewidth]{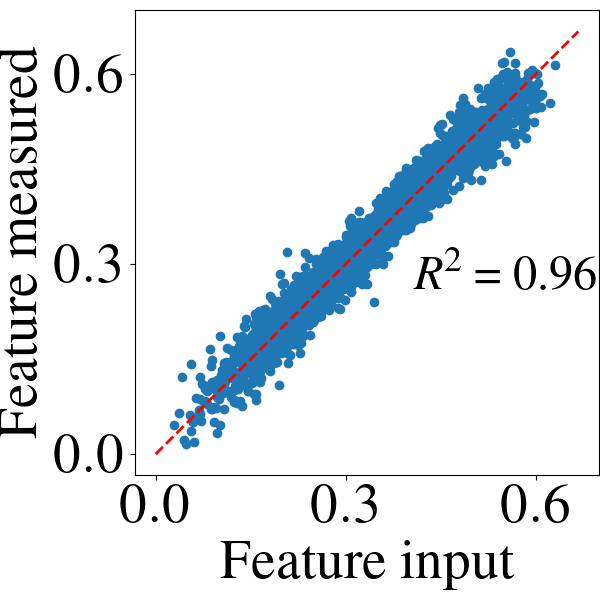}
        \caption{Phase A volume fraction}
        \label{fig:phaseA}
    \end{subfigure}
    \begin{subfigure}[b]{0.19\linewidth}
        \includegraphics[width=0.99\linewidth]{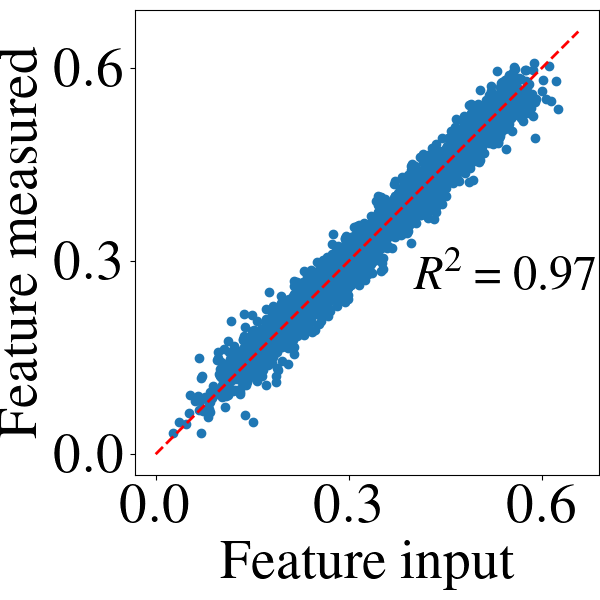}
        \caption{Phase B volume fraction}
        \label{fig:phaseB}
    \end{subfigure}
    \begin{subfigure}[b]{0.19\linewidth}
        \includegraphics[width=0.99\linewidth]{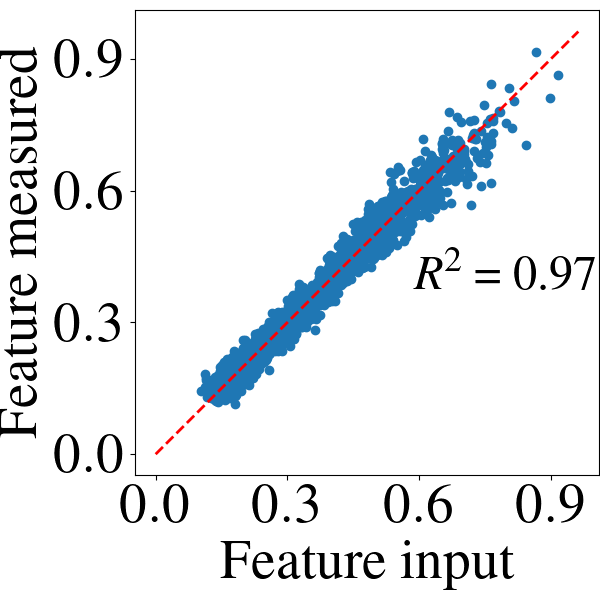}
        \caption{Mixed volume fraction}
        \label{fig:phaseMixed}
    \end{subfigure}   
    \begin{subfigure}[b]{0.19\linewidth}
        \includegraphics[width=0.99\linewidth]{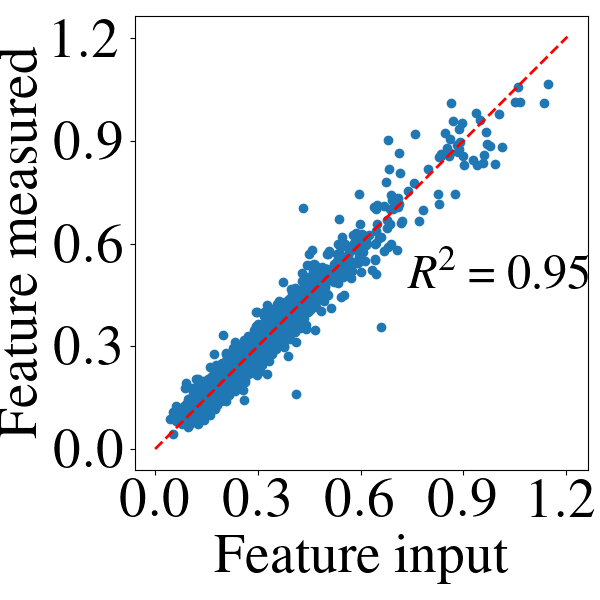}
        \caption{Tortuosity A}
        \label{fig:Tortuosity A}
    \end{subfigure}
    \begin{subfigure}[b]{0.19\linewidth}
        \includegraphics[width=0.99\linewidth]{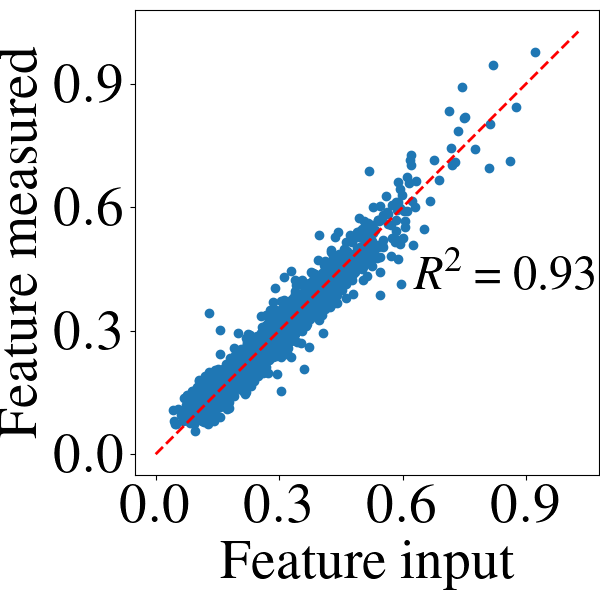}
        \caption{Tortuosity B}
        \label{fig:Tortuosity B}
    \end{subfigure}    
    \caption{Statistical analysis of conditional microstructure generation: Correlations between all features of interest, user inputs, and the corresponding features measured from generated microstructures.}
    \label{fig:conditional_generation_3p_correlation}
\end{figure}

To rigorously assess the model’s conditional generation capabilities, we produced 3,200 microstructures across a variety of targeted volume fractions and tortuosity values. We systematically compared these microstructure attributes with the user-specified conditioning parameters, as depicted in \figref{fig:conditional_generation_3p_correlation}. Our analysis reveals a high degree of accuracy in conditional generation, achieving Pearson correlation coefficients (R²) of 0.93 or greater. This robust correlation underscores the LDM’s effectiveness in adhering to precise user-defined constraints, thereby enabling targeted material design and optimization that surpasses prior methods in  versatility, and computational efficiency~\citep{hsu2021microstructure,chun2020deep}.

\subsection{Diversity and prediction of manufacturing parameters}
\begin{figure}[!h]
    \centering
    \begin{subfigure}[b]{0.55\linewidth}
        \centering
        \includegraphics[width=0.8\linewidth]{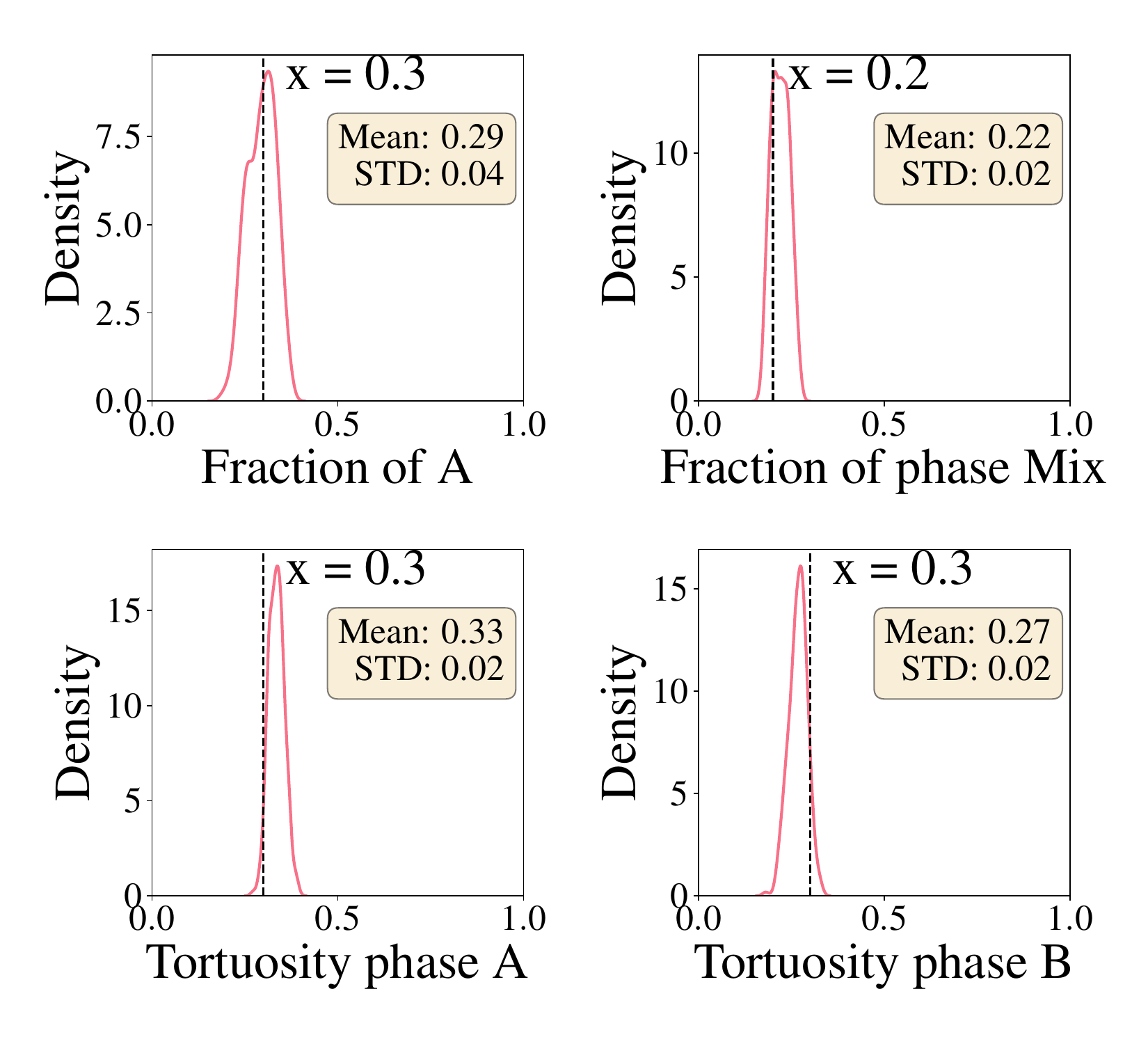}  
        \caption{Distribution of features measured from generated microstructures given specific conditional feature inputs. The vertical dotted black lines indicate the user inputs.}
        \label{fig:kde}
    \end{subfigure}
    \hfill
    \begin{subfigure}[b]{0.43\linewidth}
        \centering
        \includegraphics[width=0.99\linewidth]{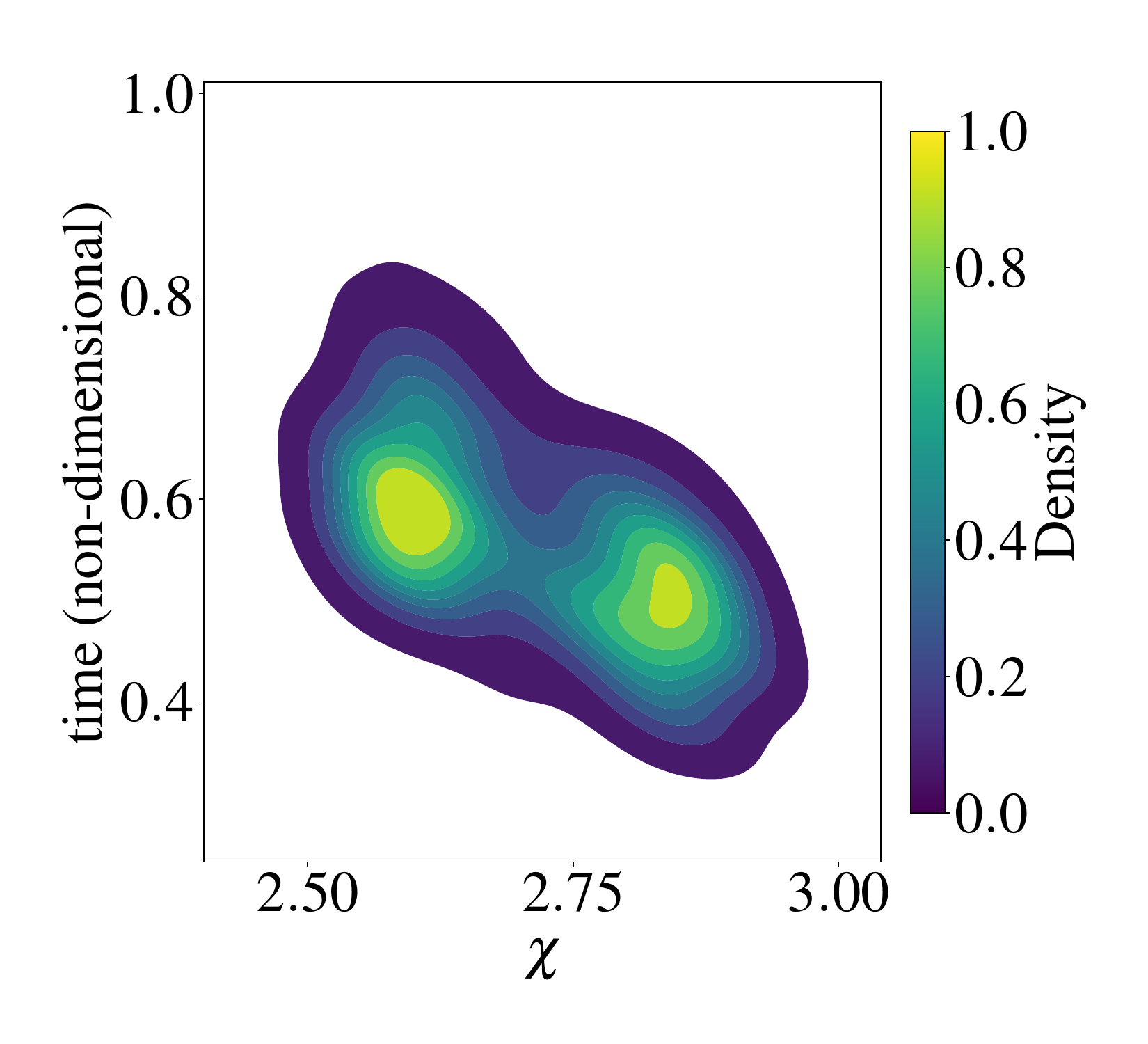}
        \caption{Contour plot of manufacturing parameters \(\chi\)and timesteps for desired microstructure generation.}
        \label{fig:contour}
    \end{subfigure}
    \caption{Variety of microstructures generated by the LDM given identical user inputs. The model can also suggests the manufacturing conditions required to generate such microstructures.}
    \label{fig:diverge_morph}
\end{figure}

We further assessed the LDM’s capability to generate diverse microstructures from identical conditional inputs. Specifically, we sampled 3200 microstructures using consistent input parameters (volume fractions: 0.3 for phase A, 0.2 for the mixed phase; tortuosities for both phases: 0.3). The resulting microstructures, detailed in the Supporting Information, exhibit significant morphological diversity despite identical conditioning parameters. \figref{fig:kde} illustrates the distributions of the extracted microstructural features, clearly aligning with the specified input values (indicated by vertical dotted lines). The strong alignment confirms that the LDM reliably generates diverse yet precisely targeted microstructures.

Moreover, \figref{fig:contour} presents contour plots predicting the manufacturing parameters --- the blend ratio, the interaction parameter ($\chi$), and the annealing time (timesteps) --- required for realizing these microstructures. Notably, the LDM framework identifies multiple feasible fabrication pathways: a combination of higher $\chi$ values with shorter annealing durations, or lower $\chi$ values with extended annealing periods. This data-driven insight aligns well with the known physical behavior of phase-separating systems described by the Cahn–Hilliard model, where increased interaction parameters accelerate phase separation, thereby requiring less annealing time, whereas lower interaction parameters necessitate longer annealing to achieve comparable morphologies. This pathway prediction capability illustrates the integration of computational design with experimental manufacturability, thus significantly advancing current microstructure design methodologies \citep{hsu2021microstructure,chun2020deep}. Such an approach could be expanded to include other manufacturing parameters, making the model applicable across various material systems and manufacturing processes \citep{hsu2021microstructure,chun2020deep}.

\begin{figure}[!t]
    \centering
    \begin{subfigure}[b]{0.3\linewidth}
        \includegraphics[width=0.99\linewidth]{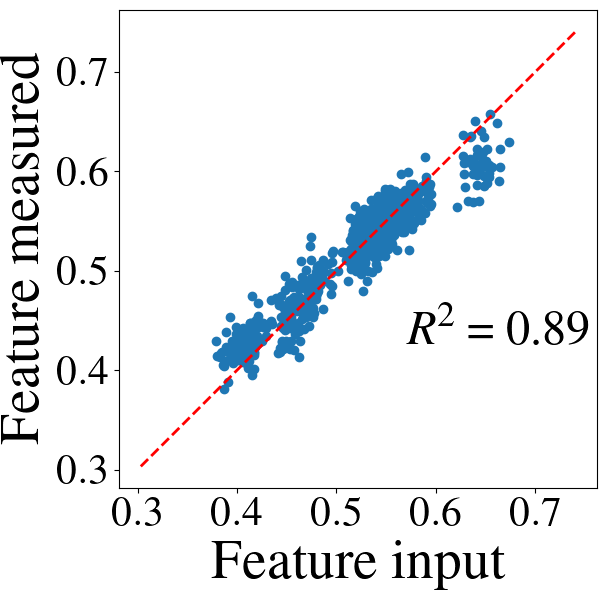}
        \caption{Donor volume fraction}
        \label{fig:exp_vol_frac}
    \end{subfigure}
    \hspace{0.03\linewidth}
    \begin{subfigure}[b]{0.3\linewidth}
        \includegraphics[width=0.99\linewidth]{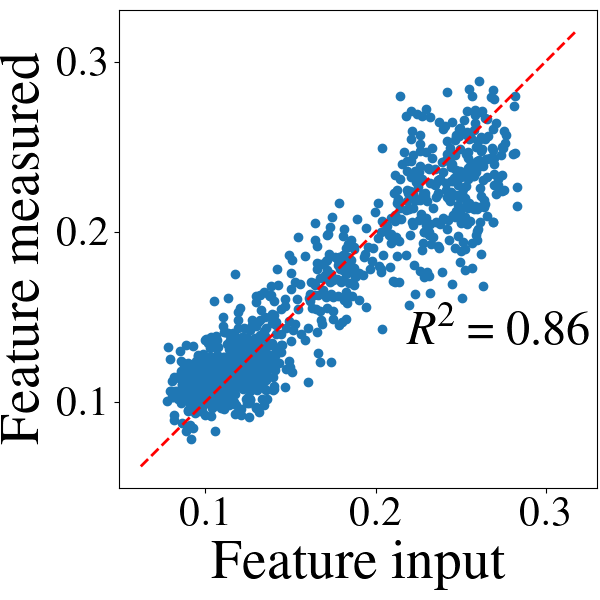}
        \caption{Tortuosity Acceptor}
        \label{fig:exp_tortA}
    \end{subfigure}
    \hspace{0.03\linewidth}
    \begin{subfigure}[b]{0.3\linewidth}
        \includegraphics[width=0.99\linewidth]{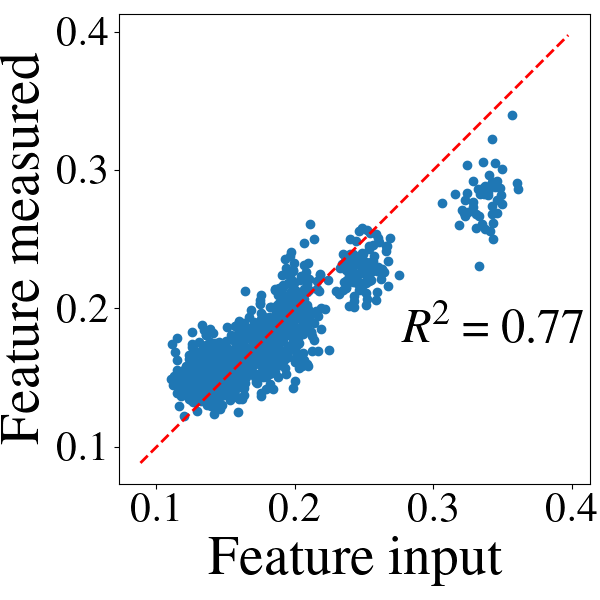}
        \caption{Tortuosity Donor}
        \label{fig:exp_tortD}
    \end{subfigure}
    \caption{Statistical analysis of conditional microstructure generation: Correlations between all features of interest, user inputs, and the corresponding features measured from generated microstructures. Unlike the synthetic dataset we observe \(R^2\) is below 0.9 }
    \label{fig:conditional_generation_experimental_correlation}
\end{figure}

\begin{figure}[!t]
    \begin{subfigure}[t]{\linewidth}
        \centering
        \includegraphics[width=0.5\linewidth]{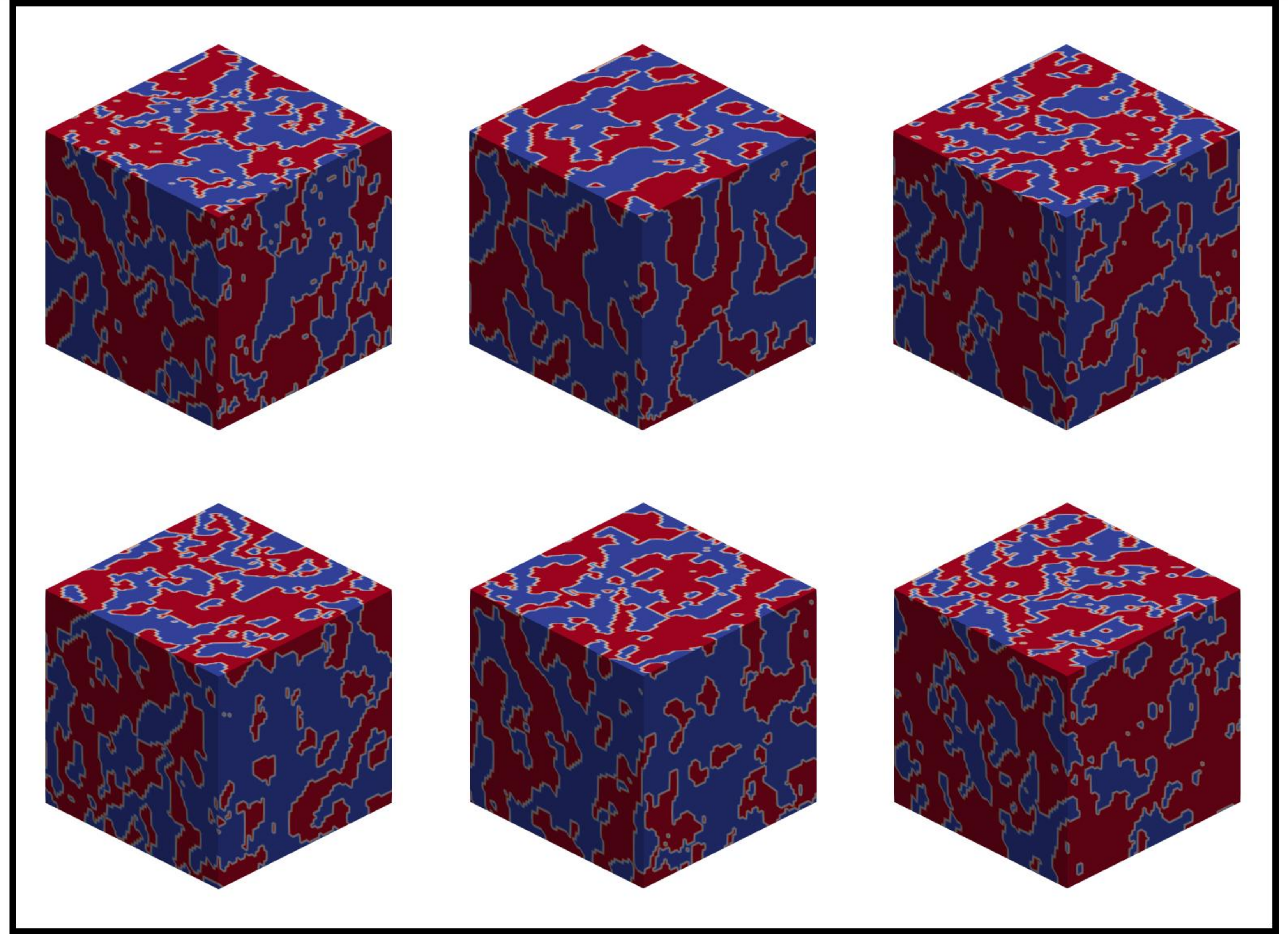}
        \caption{Samples microstructures generated from same conditional feature inputs.}
        \label{fig:exp_div_morph_samples}
    \end{subfigure}    
    \centering
    \begin{subfigure}[t]{\linewidth}
        \centering
        \includegraphics[width=0.8\linewidth]{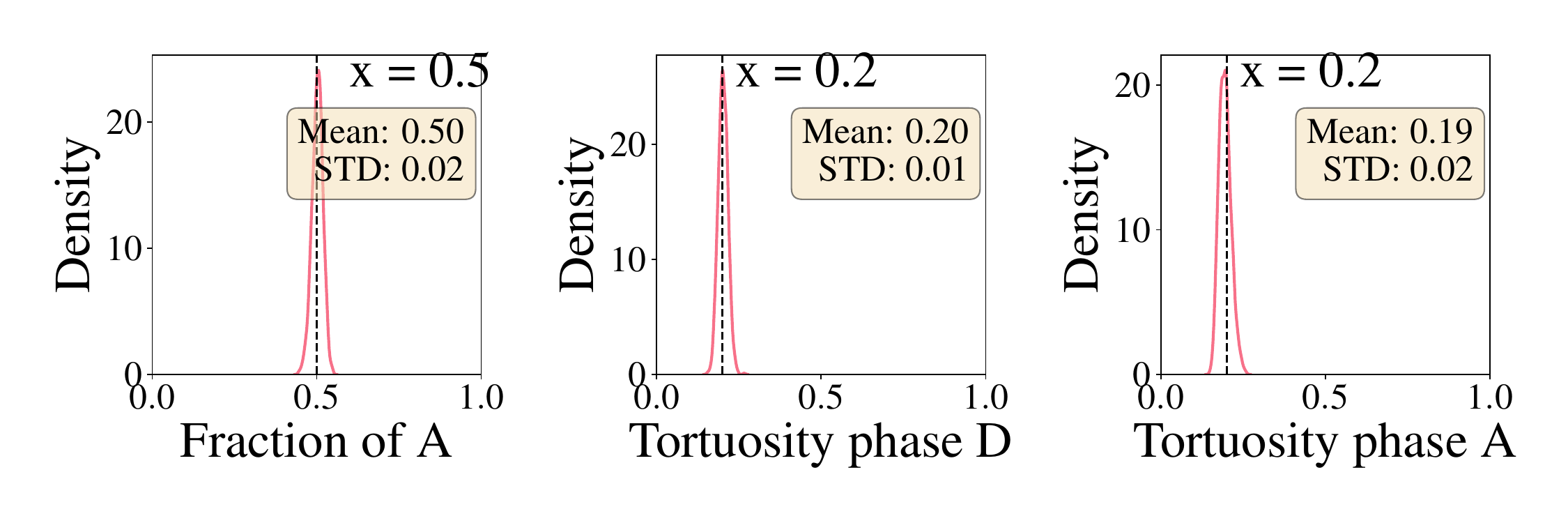}  
        \caption{Distribution of features measured from generated microstructures given specific conditional feature inputs. The vertical dotted black lines indicate the user inputs.}
        \label{fig:exp_kde}
    \end{subfigure}
    \caption{Variety of microstructures generated by the LDM given identical user inputs. The model can also suggest the manufacturing conditions required to generate such microstructures.}
    \label{fig:exp_diverge_morph}
\end{figure}

\subsection{Experimental Microstructures}
We further demonstrated our framework's applicability using an experimental dataset comprising voxelized organic photovoltaic (OPV) morphologies from spin-cast P3HT:PCBM thin films, reconstructed through tomographic energy-filtered TEM \citep{heiber2020charge,herzing20103d} (additional methodological details are provided in the Methods section).

Using this experimental dataset, we generated 1000 microstructures conditioned on user-specified inputs (volume fraction: 0.5; donor and acceptor tortuosities: 0.2 each). \figref{fig:conditional_generation_experimental_correlation} shows the correlation between the specified inputs and the measured features, achieving Pearson correlation coefficients (R²) of 0.89, 0.86, and 0.77 for volume fraction, acceptor tortuosity, and donor tortuosity, respectively. Although these correlations are somewhat lower than those obtained using synthetic datasets -- likely due to the lower resolution of the experimental data --- the model captures the volume fraction with higher accuracy, as it is a simpler global descriptor. In contrast, tortuosity, a more localized and structurally complex feature, potentially requires better resolution and poses greater modeling challenges.

Additionally, \figref{fig:exp_div_morph_samples} presents six representative microstructures generated from identical conditioning inputs, illustrating notable morphological diversity. The kernel density estimation (KDE) plots shown in \figref{fig:exp_kde} confirm that the generated feature distributions are closely centered around the specified target values, with standard deviations of 0.02 or less, highlighting the precision and robustness of the conditional LDM in practical, experimental contexts.

\section{Conclusions}
\label{sec:conclusion}
Conditional microstructure generation holds considerable potential across diverse fields, including energy storage, biomedical devices, and additive manufacturing. It enables precise control over microstructural attributes to optimize material performance, durability, and functionality. In this study, we introduced a versatile, scalable LDM-based framework for generating detailed, high-resolution 3D microstructures with remarkable efficiency. Notably, our method not only produces diverse and precise microstructures (demonstrated here for organic photovoltaics) but also effectively predicts relevant manufacturing conditions.
Our illustration using experimental OPV datasets highlights the framework's ability to accurately reflect nuanced real-world morphological complexity. By closely aligning generated microstructures with experimental observations, our approach bridges computational predictions and practical manufacturing processes, empowering targeted materials engineering and enhancing our fundamental understanding of processing–structure–property relationships.

Despite these strengths, our approach currently has limitations. The sequential training methodology, where we train the VAE, feature predictor, and latent diffusion model separately, could benefit from optimization for increased efficiency. Additionally, users must cautiously select realistic conditional parameters, as unrealistic inputs can yield impractical microstructures. Future improvements might involve streamlining or parallelizing the training pipeline and enhancing the model’s usability through intuitive interfaces, input parameter validation, or automated parameter selection to ensure robust and practical microstructure generation. Addressing these aspects will further extend the accessibility and utility of the framework to broader materials research communities.

\section{Methodology}
\label{sec:methodology}
\subsection{Training dataset}
The computational dataset used in this project was synthesized from three-dimensional simulations of the Cahn-Hilliard equation, solved using the Finite Element Method (FEM). It comprises a wide range of phase separation scenarios,  captured through simulations under varying conditions defined by two parameters: the initial volume fraction (\(\phi\)) and the Flory-Huggins interaction parameter (\(\chi\)). The Cahn-Hilliad equation represents a microstructure by modeling the spatial variation of two or three components. In our dataset, \(\phi\) is varied systematically to explore a wide spectrum of initial mixture compositions, capturing the dynamics of phase separation. The interaction parameter, \(\chi\), is another key variable in the dataset. It quantifies the degree of affinity or aversion between the mixture's components. A higher \(\chi\) value signifies a strong tendency towards phase separation due to energetically unfavorable interactions, while a lower value suggests better miscibility. By altering \(\chi\), we probe different interaction regimes, from weak to strong phase-separating tendencies. For each combination of \(\phi\) and \(\chi\), the dataset captures over 400 time-stamped snapshots of a 3D Cahn-Hilliard simulation at $128 \times 128 \times 64$ resolution, providing a detailed temporal sequence of the phase separation process. There are 67 such time series, resulting in a total of over 26,800 3D microstructures. The dataset was divided into training and validation sets, with 80\% of the data allocated to training and 20\% to validation. 

\begin{figure}[!t]
    \centering
    \includegraphics[width=0.99\linewidth]{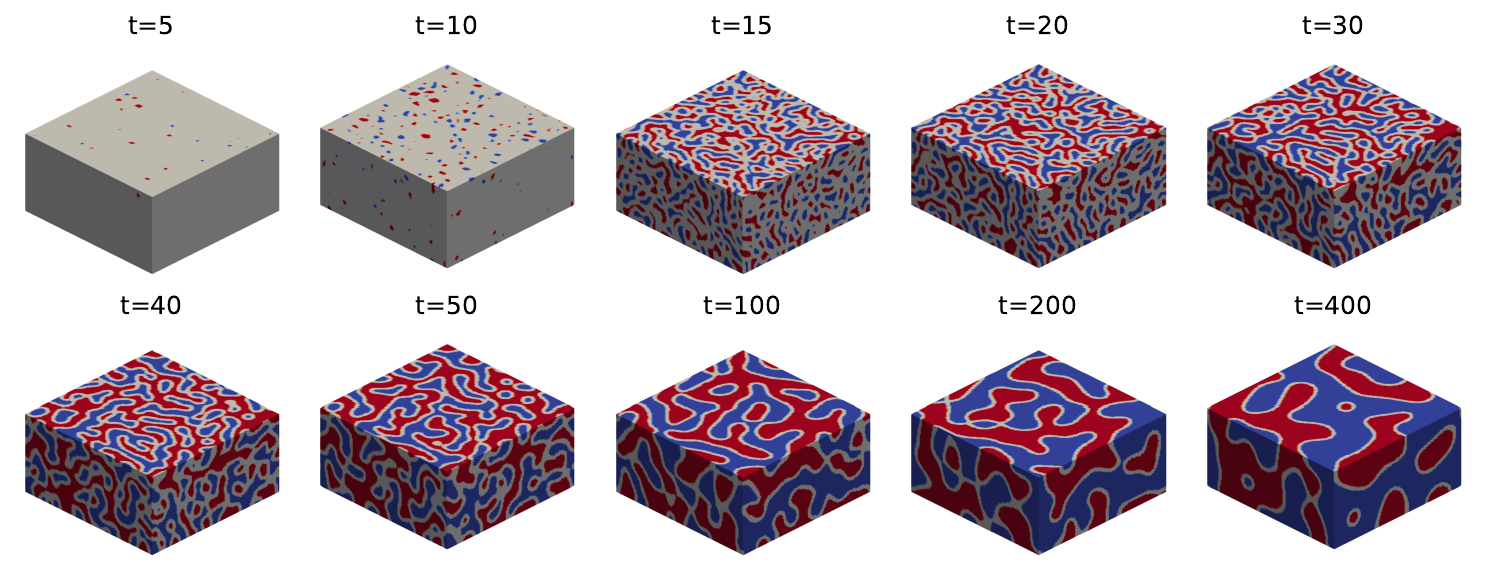}
    \caption{A sequence of 10 snapshots from one time series out of 67 in the entire dataset, illustrating the evolution of phase separation in a 3D simulation of the Cahn-Hilliard equation.}
    \label{fig:training_data_timeseries}
\end{figure}
\begin{figure}[t!]
    \centering
    \includegraphics[width=0.58\linewidth,clip,trim={2.4in 0.0in 2.4in 0.0in}]{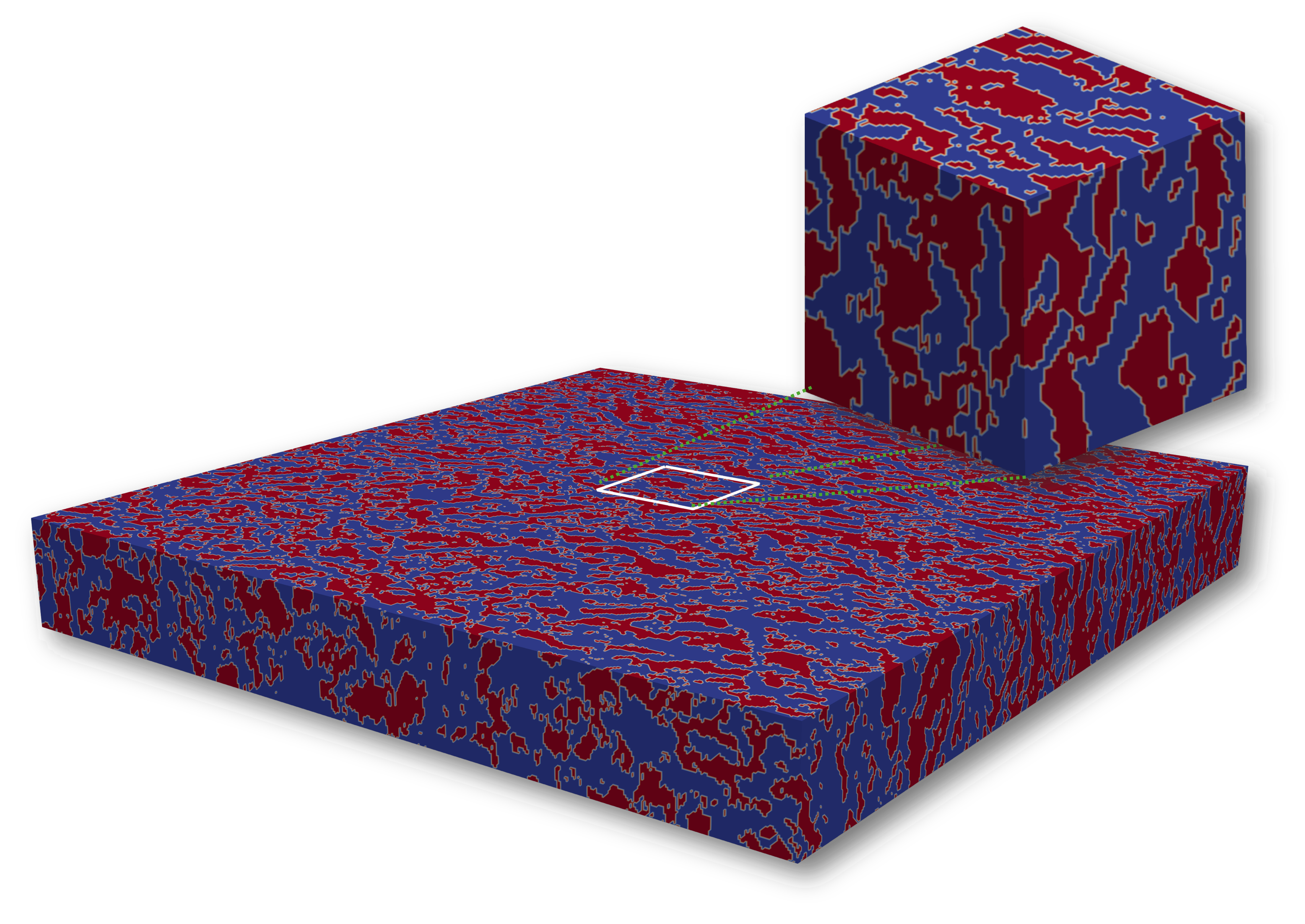}
    \caption{Visualization of spin-cast P3HT:PCBM thin film, fabricated using chlorobenzene reconstructed using tomographic energy-filtered TEM. The main image shows the reconstructured 3D morphology, with blue and red domains representing the electron-donating (donor) and electron-accepting (acceptor) materials, respectively. The inset provides a zoomed-in view of a cubic subvolume extracted from the full morphology.}
    \label{fig:morph_cb}
\end{figure}

\figref{fig:training_data_timeseries} shows snapshots from a single time series within the training dataset. The snapshots represent the temporal evolution of phase separation during the 3D simulation of the Cahn-Hilliard equation, illustrating the dynamic changes in microstructures over time. The Cahn-Hilliard model accounts for both thermodynamic forces and kinetic processes driving phase separation, providing insights into how processing conditions, such as annealing, influence the final morphology of the active layer. This understanding can aid to the optimization of material processing to improve organic solar cell (OSC) performance \citep{ronsin2022formation,konig2021two}.

In addition to the computational dataset, we also utilized voxelized experimental OPV morphologies from spin-cast P3HT:PCBM thin films fabricated using two different solvents: chlorobenzene (CB) and dichlorobenzene (DCB). These morphologies were fabricated and reconstructed using tomographic energy-filtered TEM (see \citet{heiber2020charge,herzing20103d} for details). The imaging volume had approximate dimensions of $1 \, \mu\text{m} \times 1 \, \mu\text{m} \times 100 \, \text{nm}$, with the EF-TEM-based reconstruction achieving a voxel resolution of approximately $2.12 \, \text{nm}$. The CB morphology is depicted in \figref{fig:morph_cb}, where blue domains represent the electron-donating (donor) materials and red domains indicate the electron-accepting (acceptor) materials. The voxelized resolutions of the CB and DCB morphologies are $466 \times 465 \times 50$ and $478 \times 463 \times 60$, respectively. To generate a uniform dataset, we extracted cubic subvolumes spanning the full $z$-axis of each morphology and resized them to $64 \times 64 \times 64$ using nearest-neighbor interpolation. In the $x$ and $y$ directions, we used a step size of 4 voxels, resulting in over 10,500 cubic subvolumes of size $64 \times 64 \times 64$ from each of the two main morphologies. This process yielded a total of over 21,000 $64 \times 64 \times 64$ 3D microstructures. Similar to the synthetic dataset, this dataset was also divided into training and validation sets in the usual 80\% - 20\% split.

\subsection{Generative model architecture}

\begin{figure}[t!]
    \centering
    \includegraphics[width=0.8\linewidth,clip,trim={2.4in 0.0in 2.4in 0.0in}]{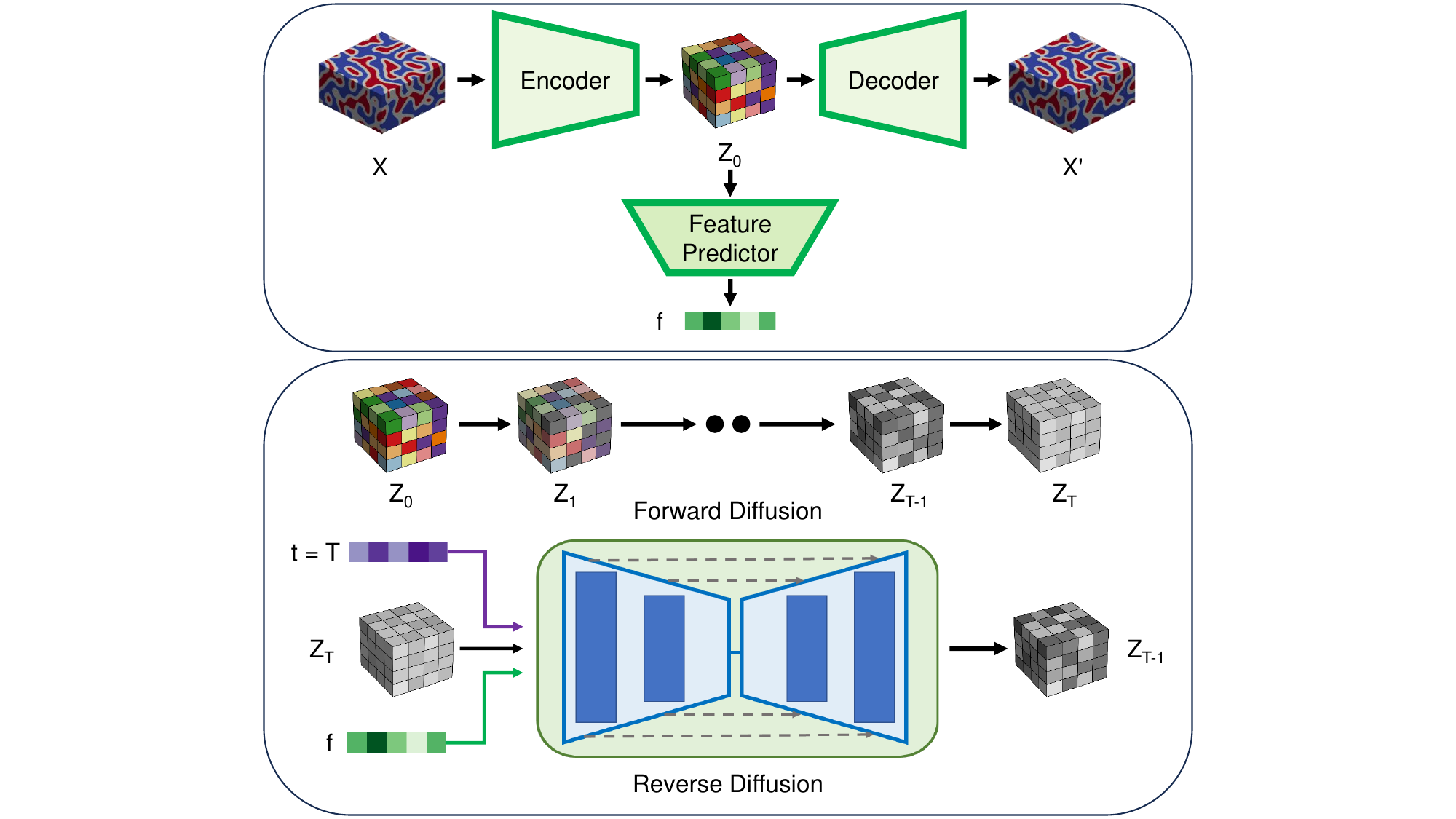}
    \caption{Overview of the proposed LDM-based framework's three-step training process: VAE training and latent representation dataset creation, training of the FP, training of DM in the latent space}
    \label{fig:Training}
\end{figure}

The architecture of the training framework is provided in \figref{fig:Training}. The core of our generative framework is the LDM, which offers several advantages over traditional DMs. LDMs are superior in computational efficiency, memory usage, generation speed, and scalability~\citep{rombach2022high,pinaya2022brain}. They excel in processing 3D data, operating in a lower-dimensional latent space that significantly reduces the computational load. This approach not only accelerates generation but also decreases memory requirements---crucial for handling complex 3D datasets. The reduced computational and memory demands allow for quicker iterations, making LDMs ideal for applications that require rapid prototyping or extensive simulations. 

Additionally, the scalability of LDMs enables them to manage larger datasets and more complex microstructures without a proportional increase in resource consumption, unlike traditional DMs. This combination of factors renders LDMs a more efficient and practical choice for generating detailed 3D microstructures in a resource-conscious manner. Our LDM framework comprises three components: a VAE, a Feature Predictor (FP), and a DM, which are trained sequentially. The encoder and decoder of the VAE are trained simultaneously to obtain the latent space from which the FP is trained. Once the VAE and FP are trained, we train the DM using the latent space and the predicted features.

\subsubsection{Variational Autoencoder}
Contrary to classic Autoencoders that transform an input x directly into a latent representation z, VAEs convert x into a probability distribution~\citep{kingma2013auto}. In VAEs, the encoder doesn't predict a single point but instead determines the mean and variance of this distribution. The latent variable z is then derived from this distribution. This is done by initially sampling from a standard normal Gaussian distribution, then scaling this sample with the predicted variance, and finally, adding the predicted mean to this scaled value. 

To generate a sample \( \mathbf{z} \) from the latent space, the VAE uses a random sample \( \epsilon \) drawn from a standard normal distribution:
\begin{equation}
\mathbf{z} = \mu_\phi(\mathbf{x}) + \sigma_\phi(\mathbf{x}) \odot \epsilon, \quad \epsilon \sim \mathcal{N}(0, \mathbf{I})
\label{eq:z}
\end{equation}
where \( \odot \) denotes element-wise multiplication. The encoder maps the input \( \mathbf{x} \) to two parameters in the latent space - the mean \( \mu \) and the log-variance (log-var):
\begin{equation}
q_\phi(\mathbf{z} | \mathbf{x}) = \mathcal{N}(\mathbf{z}; \mu_\phi(\mathbf{x}), \exp(\text{log-var}_\phi(\mathbf{x})))
\label{eq:q}
\end{equation}
The decoder maps the latent representation \( \mathbf{z} \) back to the input space:
\begin{equation}
p_\theta(\mathbf{x} | \mathbf{z}) = \mathcal{N}(\mathbf{x}; \mu_\theta(\mathbf{z}), \exp(\text{log-var}_\theta(\mathbf{z})))
\label{eq:p}
\end{equation}
The loss function in VAEs consists of two terms, the reconstruction loss and the KL divergence:
\begin{equation}
\mathcal{L}(\theta, \phi; \mathbf{x}) = - \mathbb{E}_{q_\phi(\mathbf{z} | \mathbf{x})} \left[ \log p_\theta(\mathbf{x} | \mathbf{z}) \right] + \text{KL}(q_\phi(\mathbf{z} | \mathbf{x}) || p(\mathbf{z}))
\label{eq:vae_loss}
\end{equation}
This function balances the accuracy of reconstruction with the regularization of the latent space.

The VAE is the entry point for our architecture. The VAE employed in this work consists of an encoder-decoder structure with residual blocks for feature extraction and reconstruction. The encoder comprises five 3D convolutional layers, each followed by Instance Normalization and a residual block to capture spatial dependencies in the input data. The latent space is parameterized by a mean (`mu`) and log-variance (`logvar`), both of which are obtained through additional 3D convolutional layers. The decoder mirrors the encoder's structure, using transposed convolutions to upsample the latent space back to the original input dimensions with residual blocks and Instance Normalization for stable training. A final Sigmoid activation is applied to the output to generate the reconstructed data. Once the VAE is trained, we use its encoder to compress microstructures with over a million voxels into a compact encoded representation of size 1024 ($4 \times 8 \times 8 \times 4$), while for experimental VAE inputs of $64 \times 64 \times 64$ (over 262K voxels), the output is further reduced to 512 ($1 \times 8 \times 8 \times 8$). This reduced-dimensional latent space, distinguished by its efficiently learned data distribution, facilitates more efficient and stable diffusion processes.

\subsubsection{Feature predictor}
The feature predictor is a fully connected neural network designed to predict specific microstructural and manufacturing features based on encoded representations of 3D morphological data. The model architecture includes an input layer, two hidden layers, and an output layer. The input layer receives a flattened latent representation of size 1024, generated by a pretrained VAE. This representation is then processed through two hidden layers, each reducing the data dimensionality while applying Instance Normalization and Dropout (dropout=0.1) to prevent overfitting. The final output layer maps the processed data to the desired number of features, which correspond to the predicted manufacturing and morphological characteristics.

\subsubsection{Diffusion model}
DMs consist of two main stages: the forward diffusion and the backward diffusion. In the forward diffusion stage, Gaussian noise is repeatedly added to a data sample drawn from a specific target distribution. This process is performed multiple times, resulting in a series of samples that become increasingly noisy compared to the original data. This process is described by the Markov chain:
\begin{equation}
    q(x_t \mid x_{t-1}) = \mathcal{N}(x_t; \sqrt{1 - \beta_t} x_{t-1}, \beta_t \mathbf{I})
    \label{eq:forward_diffusion}
\end{equation}
where $x_0$ is the initial sample from the target distribution $q(x)$, and the variance schedule is defined as $\{\beta_t \in (0, 1)\}_{t=1}^T$. Conversely, the backward diffusion stage aims to iteratively eliminate the noise introduced in the forward stage, represented as $q(x_{t-1} \mid x_t)$. Direct sampling from $q(x_{t-1} \mid x_t)$ is not possible because that would require the complete knowledge of the distribution. Therefore, the model uses a neural network \( G_\theta(x_{t-1} \mid x_t) \), parameterized by \( G \) and \( \theta \), to approximate these conditional probabilities. The network, refined through gradient-based optimization, aims to replicate the random Gaussian noise used in the forward diffusion process for transforming the original sample into a noisy version \( x_t \) at a particular timestep. The objective function is expressed as: 
\begin{equation}
    \lVert z - G_\theta(x_t, t) \rVert^2 = \lVert z - G_\theta(\sqrt{\bar{\alpha}_t} x_0 + \sqrt{1 - \bar{\alpha}_t} z, t) \rVert^2
    \label{eq:backward_diffusion}
\end{equation}
Here, \( \alpha_t = 1 - \beta_t \), \( \bar{\alpha}_t = \prod_{s=1}^{t} \alpha_s \), and \( z \sim \mathcal{N}(0, \mathbf{I}) \).

The neural network's primary role in a DM is to learn the inverse of the noise addition process. By systematically removing the noise added during the forward diffusion process, the network reconstructs the original data from its noisier versions. This process enables the generation of new, high-quality samples from completely random Gaussian noise.

In the context of enhancing the generative capabilities of DMs, incorporating a conditional vector provides a strategic augmentation of the model's architecture. By embedding conditional vector, \(c\), within both the embedding and decoder layers of the U-Net structure in the diffusion process, the model gains an additional layer of contextual guidance. This integration is mathematically articulated as \(\lVert z - G_\theta(x_t, t, c) \rVert^2 = \lVert z - G_\theta(\sqrt{\bar{\alpha}_t} x_0 + \sqrt{1 - \bar{\alpha}_t} z, t, c) \rVert^2\), where the conditional vector \(c\) is seamlessly intertwined with the noise prediction and denoising functions of the generative model, \(G_\theta\). Such an approach leverages the conditionality to steer the generative process, thereby imbuing the model with enhanced directional specificity and adaptiveness in its generation capabilities, aligning closely with the encoded conditions in \(c\).

Our LDM model operates under a linear beta schedule, which dictates the noise addition and removal process across the diffusion stages. This schedule is precomputed and stored as buffers, allowing for consistent noise manipulation during both training and sampling phases. The diffusion process involves progressively adding noise to the latent features and then denoising them through a series of timesteps to generate the final microstructure.

To guide the diffusion process, the model employs two key embedding networks:

\begin{itemize}
    \item \textbf{Time Embedding}: This network converts the current timestep into an embedding, providing temporal guidance during the denoising phase.
    \item \textbf{Context Embedding}: The context embedding network incorporates manufacturing features that condition the generation process, ensuring that the generated microstructures adhere to specific manufacturing parameters.
\end{itemize}

During the forward pass, the input 3D data is first encoded through the VAE to extract latent features. These features are then processed by a feature predictor model to obtain context features, specifically the initial four manufacturing features (e.g., two volume fractions and two tortuosities). These latent features are progressively diffused using the predefined beta schedule, with the U-Net model performing denoising at each timestep. The denoising process is informed by both time and context embeddings, enabling precise reconstruction of the microstructure. For new sample generation, the diffusion process is reversed, starting from pure noise and progressively refining the latent space into a structured representation conditioned on the context features.

\begin{figure}[b!]
    \centering
    \includegraphics[width=0.99\linewidth]{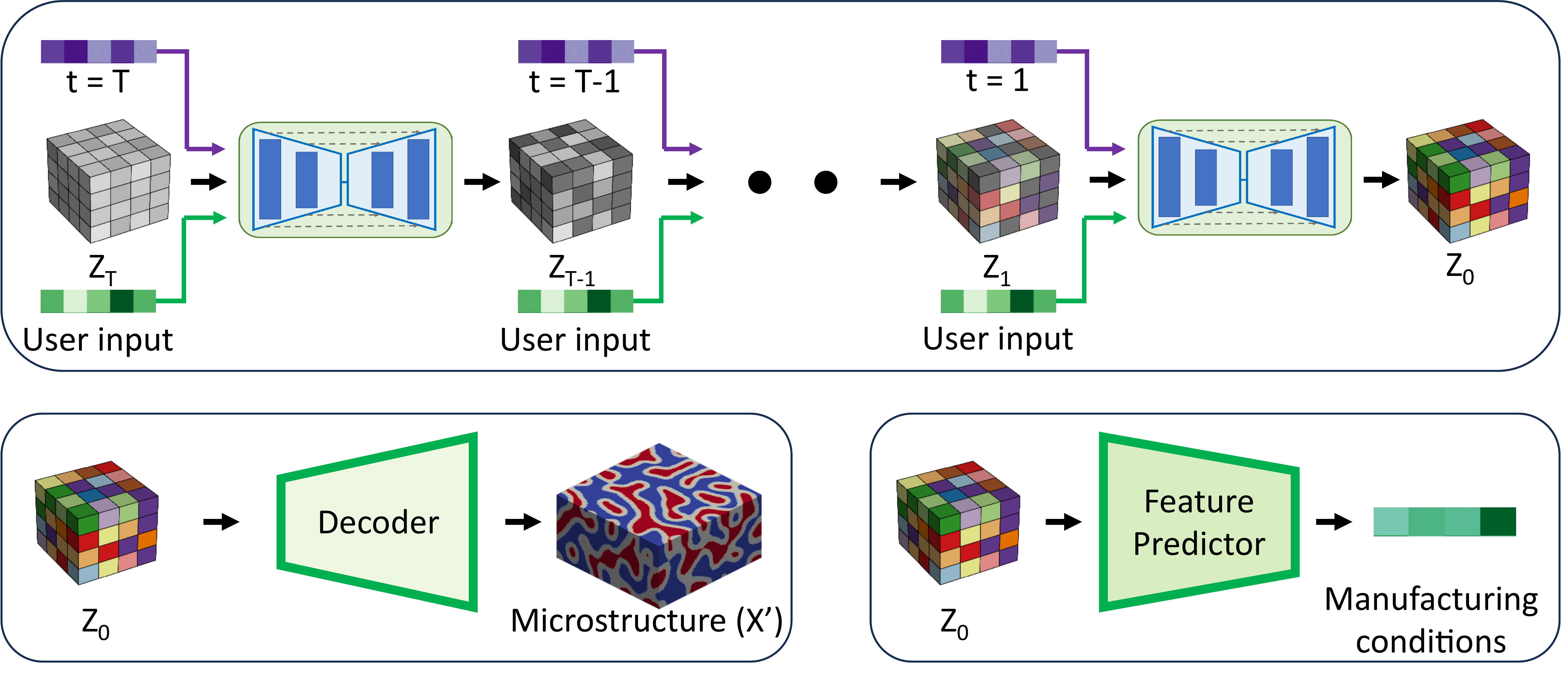}
    \caption{Overview of the inference framework for the proposed LDM-based model: Random noise \(Z_T\) is sampled in latent space, and the diffusion model gradually denoises it over \(T\) steps. User inputs condition the denoising process. \(Z_T\) is then passed through the VAE decoder and the feature predictor to obtain the microstructure and its manufacturing parameters, respectively.}
    \label{fig:inference}
\end{figure}

\subsection{Training and inference}
As shown in \figref{fig:Training}, the training process consists of three steps. First, the VAE is trained on the original training dataset. Once the VAE is trained, we encode the entire training dataset to obtain the latent representation, which becomes the training data for both the feature predictor and the diffusion model. In the second step, we train the feature predictor. Once trained, the input to the feature predictor is a latent representation of the microstructure, and the output is the features of interest, such as manufacturing parameters, tortuosity, volume fraction, etc. Finally, the LDM is trained to denoise and recover the original data from noisy inputs, with the corresponding features of interest used as conditioning. The detail of the training process is provided in the appendix.

The inference process begins with the pre-trained weights of the LDM, VAE decoder, and feature predictor. The VAE encoder is not required during inference. The process involves user input and random noise sampled in the latent space. The random noise is iteratively refined by the LDM, conditioned on the user inputs. After 1,000 iterations, the denoised latent representation of the microstructure is obtained. This step is the most time-consuming during inference. However, despite this many iterations, the process remains highly efficient because the denoising occurs in latent space rather than pixel space, which has 1,000 times fewer dimensions. The inference pipeline is demonstrated in \figref{fig:inference}. Once the denoised latent representation of the microstructure is obtained, it is passed through both the feature predictor and the VAE decoder. The feature predictor provides the manufacturing conditions, while the VAE decoder generates the final conditioned microstructure. Using NVIDIA A100 80GB GPU cards, it takes 2 seconds to infer a single microstructure.

\section*{Acknowledgments}
This work was supported by the National Science Foundation under CMMI-2053760 and DMR-2323716. We acknowledge computing support from NSF ACCESS. 

\section*{Data Availability}
The codebase and dataset used in this work will be made public upon acceptance of the paper.

\section*{Conflict of Interest}
The authors declare that they have no conflict of interest with respect to the contents of this article.



\bibliographystyle{unsrtnat}
\bibliography{bibliography}

\appendix
\newpage
\setcounter{page}{1}
\renewcommand\thefigure{A.\arabic{figure}}    
\renewcommand\thetable{A.\arabic{table}}    
\setcounter{figure}{0}
\setcounter{table}{0}

\section{Appendix}
\subsection{Training process details and hyperparameter tuning}
\label{sec:training_process}
All three components of the architecture—the VAE, feature predictor, and LDM—were trained for 500 epochs with a batch size of 32. We chose a smaller batch size to mitigate the risk of out-of-memory errors, particularly given that we are working with 3D data. The Adam optimizer~\citep{kingma2014adam} was employed for gradient-based optimization. The Adam optimizer was selected due to its widespread adoption, stability, and efficiency. The learning rate was dynamically adjusted using a cosine annealing scheduler, which effectively reduces the loss by gradually decreasing the learning rate~\citep{loshchilov2016sgdr, loshchilov2017decoupled}. Each model took 3-4 days to train, and training all three models sequentially took a total of 11 days.

The loss function for VAE combined a Mean Squared Error (MSE) loss for reconstruction and a Kullback-Leibler Divergence (KLD) loss~\citep{kullback1951information}, with a weight of $1 \times 10^{-6}$ for regularizing the latent space. The goal was to keep both the KLD and reconstruction losses in the same order of magnitude. The feature predictor was trained using an MSE loss function to assess the accuracy of predictions by measuring the difference between predicted and actual feature values. The encoder of the pretrained VAE was kept frozen during feature predictor training phase. For both the VAE and the feature predictor, the initial learning rate was set to $5 \times 10^{-5}$, with a minimum of $5 \times 10^{-7}$.

For the LDM, the diffusion process was divided into 1000 timesteps. The training objective was to minimize the MSE between the predicted noise and the actual noise added during the diffusion process. Initial and minimum learning rates are $1 \times 10^{-6}$ and $1 \times 10^{-7}$, respectively. The learning rate was selected based on the pioneering work by \citep{rombach2022high}, which demonstrated the effectiveness of using this order of magnitude in similar architectures. Both VAE and feature predictor were kept frozen during LDM training. 

The training process for all models was conducted in a GPU-enabled environment, using an NVIDIA A100 GPU with 80 GB of memory. The entire framework was implemented in PyTorch and managed by PyTorch Lightning, which handled the training loop, logging, and checkpointing. Checkpoints were automatically saved based on the validation loss, ensuring that only the best-performing models were retained. Throughout the training, real-time progress and performance metrics were continuously logged using the WandB logger, providing detailed experiment tracking and facilitating reproducibility and scalability.

\subsection{Inference microstructure samples}
\begin{figure}[h!]
    \centering
     \begin{subfigure}[b]{0.47\linewidth}
        \centering
        \includegraphics[width=0.95\linewidth]{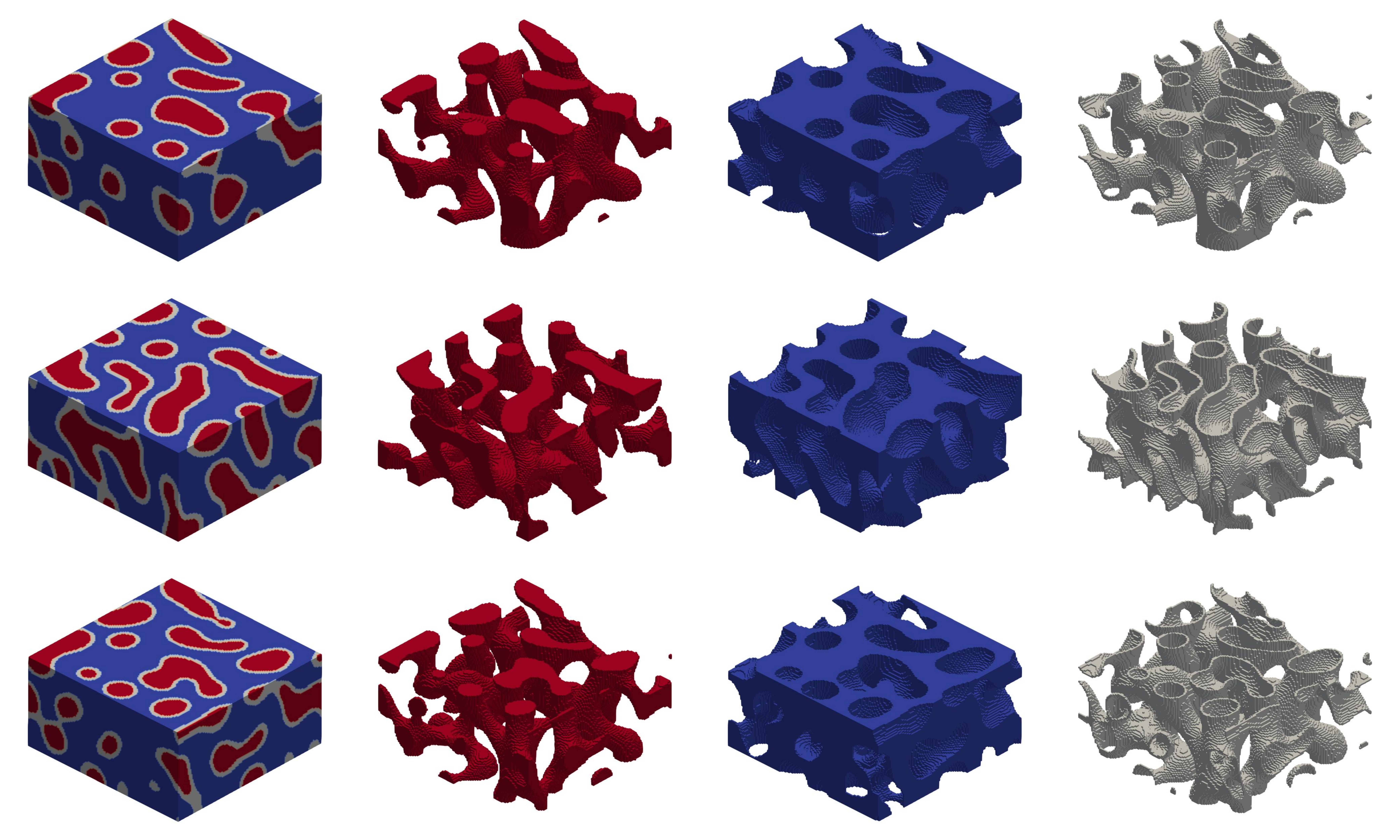}
        \caption{Sampled microstructures with a predominant phase~B (volume fraction above 0.5).}
        \label{fig:phase_B_Appendix}
    \end{subfigure}
    \hfill
    \begin{subfigure}[b]{0.5\linewidth}
        \centering
        \includegraphics[width=0.95\linewidth]{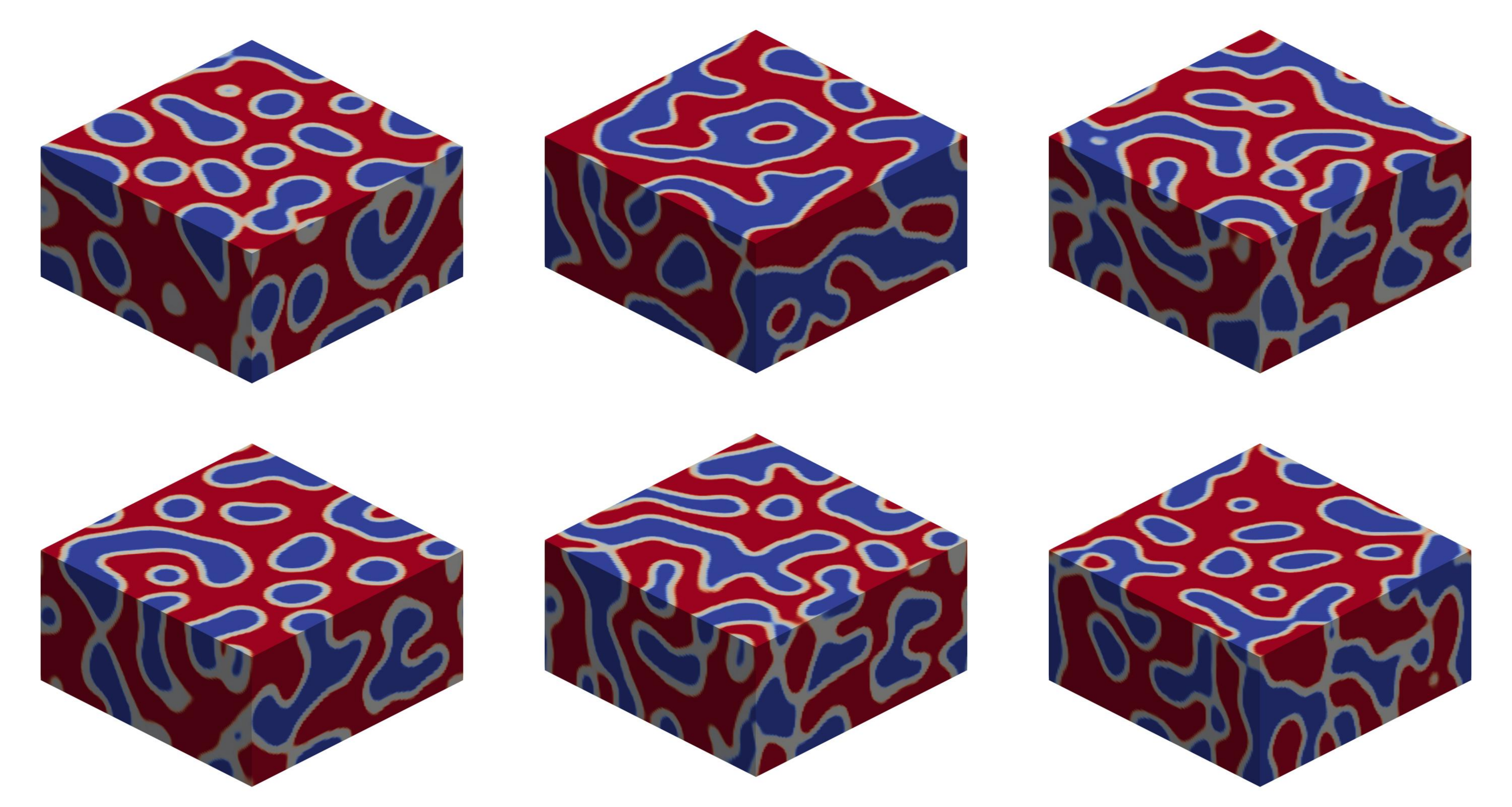}
        \caption{Microstructures generated from the same conditional features: volume fraction of phase A and phase mix 0.3 and 0.2, respectively. Tortuosity of both phases is 0.3.}
        \label{fig:diverge_morphs_samples}
    \end{subfigure}
    \caption{Three phase inference micsrostructure samples.}
    \label{fig:combined_microstructures}
\end{figure}

\clearpage

\subsection{Experimental training dataset feature distribution}
\begin{figure}[h!]
    \centering
    \begin{subfigure}[b]{0.4\linewidth}
        \centering
        \includegraphics[width=\linewidth]{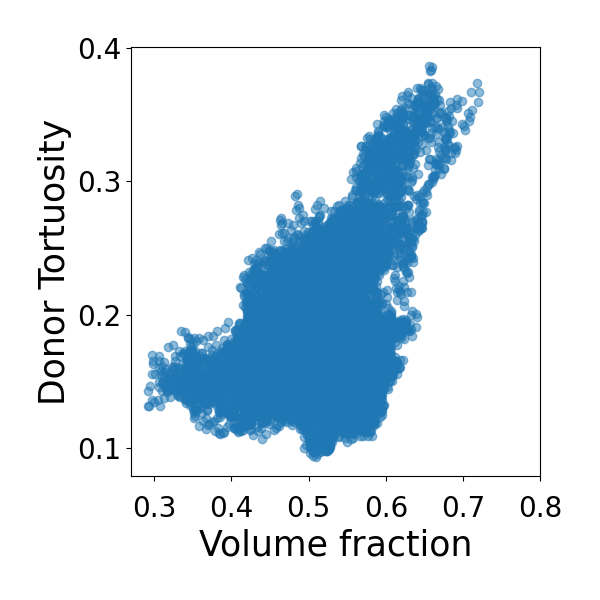}
        \caption{Donor tortuosity vs volume fraction }
        \label{fig:subfig1}
    \end{subfigure}
    \begin{subfigure}[b]{0.4\linewidth}
        \centering
        \includegraphics[width=\linewidth]{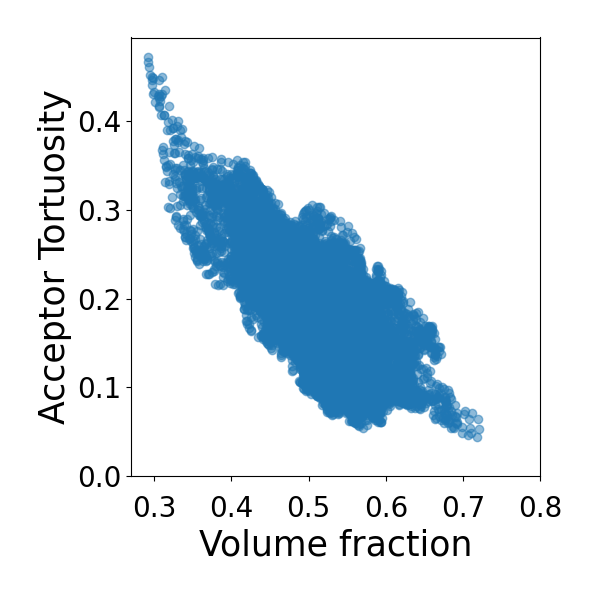}
        \caption{Acceptor tortuosity vs volume fraction}
        \label{fig:subfig2}
    \end{subfigure}
    \vspace{0.5cm}
    \begin{subfigure}[b]{0.4\linewidth}
        \centering
        \includegraphics[width=\linewidth]{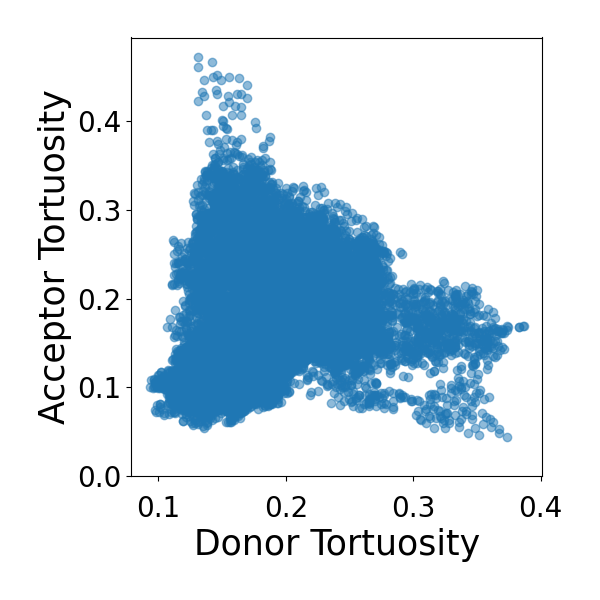}
        \caption{Acceptor tortuosity vs donor tortuosity}
        \label{fig:subfig3}
    \end{subfigure}
    \begin{subfigure}[b]{0.45\linewidth}
        \centering
        \includegraphics[width=\linewidth]{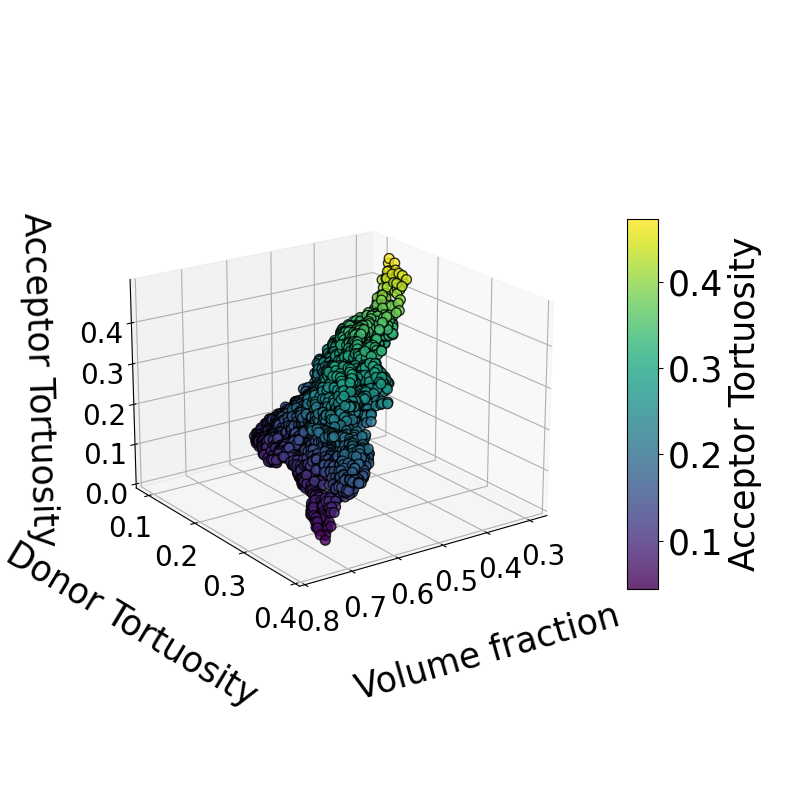}
        \caption{3D scatter plot of all three features. }
        \label{fig:subfig4}
    \end{subfigure}
    \caption{Distribution of all three features of interest. $(a)$, $(b)$, and $(c)$ show pairwise distributions, while $(d)$ presents a three-dimensional plot of all three features. This visualization highlights the range of the features, and how they are distributed relative to one another.}
    \label{fig:mainfigure}
\end{figure}

\end{document}